\documentclass[aps,showpacs,twocolumn, tightenlines]{revtex4}
\usepackage{epsfig,amssymb,bm,graphics,color}
\usepackage{epstopdf}
\newcommand*{\fs}[1]{#1\!\!\!/}

\newcommand*{\sh}{\rm sinh}

\begin{document} {\normalsize }

\title {Laser pulse-shape dependence of Compton scattering}% \\
%(2) {Multi-photon effects in Compton scattering of\\
% very short laser pulses}

\author{
%%%%%%%%%%%%%%%%%%%%%%%%%%%%%%%%%%%%%%%%%%%%%%%%%%%
 Alexander~I.~Titov$^{a,b}$,   Burkhard~K\"ampfer$^{c,d}$, Takuya~Shibata$^b$,
 Atsushi~Hosaka$^{b,e}$, and  Hideaki~Takabe$^f$}
 %%%
 \affiliation{
 $^a$Bogoliubov Laboratory of Theoretical Physics, JINR, Dubna 141980, Russia\\
 $^b$RCNP, 10-1 Mihogaoka Ibaraki, 567-0047 Osaka, Japan\\
 $^c$Helmholtz-Zentrum  Dresden-Rossendorf, 01314 Dresden, Germany\\
 $^d$Institut f\"ur Theoretische Physik, TU~Dresden, 01062 Dresden, Germany\\
 $^e$J-PARC Branch, KEK Theory Center,
 % Institute of Particle and Nuclear Studies
  KEK, Tokai, Ibaraki, 319-1106, Japan\\
 $^f$ILE, Yamada-oka, Suita, Osaka 565-0871, Japan}
%%%%%%%%%%%%%%%%%%%% Abstract %%%%%%%%%%%%%%%%%%%%%

\begin{abstract}
  Compton scattering of
  short and ultra short (sub-cycle) laser pulses off mildly relativistic
  electrons is considered within a QED framework.
  The temporal shape of the pulse
  is essential for the differential cross section as a function
  of the energy of the scattered photon
  at fixed observation angle.
  The partly integrated cross section is sensitive to
  the non-linear dynamics resulting in a large enhancement
  of the cross section for short and, in particular,
  for ultra-short flat-top pulse envelopes which can reach
  several orders of magnitude, as compared with
  the case of a long
  pulse. Such effects can be studied experimentally
  and must be taken into account in
  Monte-Carlo/transport simulations of %$e^+e^-$ pair production in
  the interaction of electrons and photons in a strong laser field.
\end{abstract}


 \maketitle

\section{introduction}

 The rapidly progressing laser technology \cite{Tajima}
 offers unprecedented opportunities for investigations
 of quantum systems with intense laser beams~\cite{Piazza}.
 A laser intensity of $\sim 2\times 10^{22}$  W/cm${}^2$ has been already
 achieved~\cite{I-22}. Intensities of the order of
 $I_L \sim 10^{23}...10^{25}$ W/cm$^2$ are envisaged in near future, e.g.\
at the CLF~\cite{CLF}, ELI~\cite{ELI}, and HiPER~\cite{hiper} laser
 facilities. The high intensities are provided in short
 pulses on a femtosecond pulse duration level~\cite{Piazza,ShortPulse},
 with only a few oscillations of the electromagnetic (e.m.) field
 or even sub-cycle pulses.
(The tight connection of high intensity and short pulse duration
 is further emphasized in \cite{Mackenroth-2011}. The attosecond
 regime will become accessible at shorter wavelengths \cite{atto}).
 These conditions are relevant for the formation of positrons
 from cascade processes in a photon-electron-positron
 plasma~\cite{Fedotov-2010,Elkina-2011} generated by
 photon-laser~\cite{Ilderton-2011},
 electron-laser~\cite{Ilderton-2010} or laser-laser
 interactions~\cite{Kirk-2009} (see \cite{Bulanov-2011} for surveys).
 The evaluation of corresponding transport equations needs as an input
 the probabilities/cross sections for the formation of $e^+e^-$ pairs
 (e.g., in the non-linear Breit-Wheeler process)
 and photons (e.g., in the non-linear Compton scattering).

 The solutions, obtained by Reiss, Nikoshov, Ritus, Narozhny
 and collaborators in a compact
 form, are valid for an infinite (both in time and space)
e.m.\ background field
 \cite{Reiss,NikishovRitus,NNR,Ritus-79,LL4}.
 That means the infinite pulse approximation (IPA) is
 used often in the above studies.
 However, the analysis of the non-linear Breit-Wheeler processes in a short
 plane-wave pulse (we call this the finite pulse approximation (FPA))
 performed recently~\cite{TKTH-2013,TTKH-2012,Nousch,Krajewska}
 shows striking differences
 between results of IPA and FPA which can reach several orders of magnitude
 depending on the pulse shape, duration, and intensity.
 Obviously, such significant effects must be properly taken into account
 in corresponding simulations for the development of seeded cascades~\cite{King}.
 It seems therefore naturally to perform
 a similar analysis for the Compton scattering,
 thus further developing recent approaches in
 \cite{Boca-2009,Heinzl-2009,Mackenroth-2011,Seipt-2011,Dinu,Seipt-2012,Krajewska-2012},
 where FPA effects have been addressed.
 Most of these papers were focused on fully differential cross
 sections with relatively simple one-parameter envelope functions,
 e.g. $\cos^2$, Gauss, $\sh$, box or related shapes.
 It was found that the fully
 differential cross section has a complicated structure
 being a rapidly oscillating function of the frequency of the
 outgoing photons $\omega'$ at fixed scattering angle $\theta'$,
 especially in  the most  interesting
 kinematically forbidden region for one-photon emission
 enabled by the multi-photon dynamics.
 The spectral structure of the cross section is rather involved
 and determined by the properties
 of the pulse structure, namely its shape and duration,
 as well as field intensity and kinematics.
 Since the Compton scattering in the multi-photon (cumulative) region
 has a basic and applied significance as a source of hard photons
it should be interesting and important to answer the following
 questions
 (i) what  observables and
 kinematical conditions are preferable for a manifestation of
 the non-linear dynamics, which may be related to the multi-photon
 effects,
 (ii) what is  role of the pulse structure
 (shape and duration), and
 (iii) under which conditions
 the predictions of FPA are close to that of IPA
 which is important for the design of transport approaches.
 The aim of our paper is to clarify these questions.

 For the sake of completeness, we start our analysis
 from fully differential cross sections which
 are calculated as a  function
 of the frequency of the outgoing photon at fixed scattering angle.
 The main difference to the previous studies
 mentioned above is utilizing a wider class of
 the pulse envelope functions including flat-top envelopes.
 Also here, it is shown that the fully
 differential cross section has a complicated structure
 being rapidly oscillating function of
% the energy of the outgoing photons
$\omega'$ at fixed $\theta'$,
 especially in the kinematically
 forbidden region.
% determined by the multi-photon dynamics.
 It is clear that
 experimentally studying the multi-photon dynamics in case of rapidly
 varying cross sections is a challenging task.
 Rather integrated observables may overcome
 this problem.

 But here one has to be careful. The totally
 integrated cross section is not suitable for this aim, since
 in this case the integration starts from the minimum value of
 the energy of the outgoing photon, $\omega'_1$,
 kinetically allowed for the one-photon emission process, and this region dominates
 in the total cross section, masking the relatively weak effects
 of multi-photon interactions. To highlight the role of the
 multi-photon interaction the lower limit of integration $\omega'$ must be
 shifted relative to $\omega'_1$:  $\omega'>\omega_1'$.
 Such partly integrated cross sections
 are smooth functions of $\omega'$ and allow to study directly the multi-photon
 dynamics enabling a clarification of the
 items (i-iii) formulated above.
 Going this way we elaborate a method for the
 calculation of the cross section of Compton
 scattering in the non-linear (multi-photon) regime
 accounting for the effect of the finite laser pulse duration with
 emphasis on different temporal pulse shapes.
 Our analysis is based on methods developed in \cite{TKTH-2013}
 for the non-linear Breit-Wheeler processes which is a crossing channel of the
 non-linear Compton effect. Despite of the similarities
 between the two processes the physical meaning of
 the dynamical variables and observables are quite different.

 Our paper is organized as follows.
 In Sect.~II we derive  the basic expressions  for the relevant observables
 in Compton scattering in FPA. In Sect.~III we discuss results of
 numerical calculations. Our summary is given in Sect.~IV.

\section{General formalism}

The Compton process is considered here as the spontaneous emission
of one photon off an electron in an external e.m.\ wave. We employ
the four-potential of a circularly polarized laser field in
the axial gauge $A^\mu=(0,\,\vec A(\phi))$ with
\begin{eqnarray}
 \vec{A}(\phi)=f(\phi) \left( \vec a_1\cos(\phi+\tilde\phi)+ \vec
 a_2\sin(\phi+\tilde\phi)\right)~, \label{III1}
 \end{eqnarray}
 where $\phi=k\cdot x$ is invariant phase with four-wave vector
 $k=(\omega, \vec k)$, obeying the null field property $k^2=k\cdot
 k=0$ (a dot between four-vectors indicates the Lorentz scalar
 product) implying $\omega = \vert \vec k \vert$, $ \vec a_{(1,2)} \equiv \vec a_{(x,y)}$;
 $|\vec a_x|^2=|\vec a_y|^2 = a^2$, $\vec a_x \vec a_y=0$;
 transversality means $\vec k \vec a_{x,y}=0$ in the present gauge.
 The envelope function $f(\phi)$ with
 $\lim\limits_{\phi\to\pm\infty}f(\phi)=0$ (FPA) accounts for the
 finite pulse length. (IPA would mean $f(\phi) = 1$).
 To define the pulse duration one can use the number $N$ of cycles in a
 pulse, $N=\Delta/\pi=\frac12\tau\omega$, where the dimensionless
 quantity $\Delta$ or the duration of the pulse $\tau$ are further
 useful measures.
 The carrier envelope phase $\tilde\phi$ is particularly important
 if it is varied in a range comparable with the pulse duration
 $\Delta$. In IPA it is
 anyhow irrelevant; in FPA with $\tilde\phi\simeq\Delta$ the cross
 section of the photon emission would be determined by an involved
 interplay of the carrier phase, the pulse duration and pulse shape
 as well as the intensity of e.m. field  as emphasized, e.g., in
 \cite{Mackenroth}) (see also \cite{Krajewska,Hebenstreit}).
 In present work, we drop the carrier phase, thus assuming
 $\tilde\phi\ll\Delta$, and concentrate on the dependence of the
 cross sections on the parameters responsible essentially for
 multi-photon effects. A detailed analysis
 of the impact of $\tilde\phi$ on the photon emission needs a
 separate investigation which is postponed to subsequent work.
 We also drop a consideration of the pulse focusing (which,
 however, is more relevant for longer pulses~\cite{NF1996}
 than those covered in the present paper) leaving such an analysis
 for forthcoming works.
%%%%%%%%

 Below, we are going to analyze the dependence of observables on
 the shape of $f(\phi)$ for two types of envelopes:
 the one-parameter hyperbolic secant~(hs) shape and the two-parameter
 symmetrized Fermi~(sF) shape
 \begin{eqnarray}
 f_{\rm hs}(\phi)=\frac{1}{\cosh\frac{\phi}{\Delta}}~,\qquad
 f_{\rm sF}(\phi)=\frac{\cosh\frac{\Delta}{b} +1}
 {\cosh\frac{\Delta}{b} +\cosh{\frac{\phi}{b}}}~.
 \label{E1}
 \end{eqnarray}
 These two shapes cover
 a variety of relevant envelopes discussed in literature
 (for details see \cite{TKTH-2013}).
 The parameter $b$ in the sF shape describes the ramping time
 in the neighborhood of $\phi \sim \Delta$. Small ratios $b/\Delta$
 cause a flat-top shaping. At $b/\Delta\to0$, the sF shape
 becomes a rectangular pulse~\cite{Boca-2009}.
 In the following, we choose the ratio
 $b/\Delta$ as the second independent
 parameter for the sF envelope function.

 The intensity of the e.m.\ field is described by the dimensionless parameter
 $\xi^2=\frac{e^2 a^2}{m^2}$, where $m$ is the electron mass
 (we use natural units with $c=\hbar=1$, $e^2/4\pi = \alpha \approx
 1/137.036$). A second relevant variable is the total energy in an
 electron--one-photon interaction $s = m^2+2(E + |\vec p|)\,\omega $,
 where $E$ and $\omega$ are the electron energy and the
laser background-field photon-frequency in the laboratory system,
and we consider head-on collisions.

Using the e.m.\ potential (\ref{III1}) and the Volkov solution for
the electron wave function in that field leads to the following
expression for the $S$ matrix element
\begin{eqnarray}
S =-i e \int\limits_{-\infty}^\infty dl \, M(l) \frac{(2\pi)^4
\delta^4(p+ lk-p'-k')}{\sqrt{2E \, 2E' \, 2\omega'}}~,
 \label{ES}
\end{eqnarray}
 where $k$, $k' = (\omega', \vec k')$, $p = (E, \vec p)$ and $p' = (E', \vec p')$
 refer to the four-momenta of the
 background (laser) field (\ref{III1}), scattered photon,
 as well as asymptotic incoming (in-state)
 and outgoing (out-state) electrons in the Furry picture.
 All quantities are considered in the laboratory system. The
 transition matrix $M(l)$ consists of four terms
 (cf.~\cite{Boca-2009}),
 \begin{eqnarray}
 M(l)=\sum\limits_{i=0}^3  M^{(i)}\,C^{(i)}(l)~,
 \label{EM}
 \end{eqnarray}
 where the transition operators have the form
 $M^{(i)}=\bar u_{p'}\,\hat M^{(i)}\,u_p$ % \label{B1}
 with
\begin{eqnarray}
 \hat M^{(0)}&=&\fs\varepsilon'~,\quad
 \hat M^{(1)}=
  \frac{ e^2a^2 \,
 (\varepsilon'\cdot k)\,\fs k}
 {2(k\cdot p)(k\cdot p')}~,\nonumber\\
 \hat M^{(2,3)}&=&\frac{e\fs a_{(1,2)}\fs k\fs
 \varepsilon'}{2(k\cdot p')} + \frac{e\fs \varepsilon'\fs k\fs
 a_{(1,2)}}{2(k\cdot p)}~.
 %\qquad \hat M^{(3)}=\frac{e\fs a_2\fs k\fs
 %\varepsilon'}{2(k\cdot p)} + \frac{e\fs \varepsilon'\fs k\fs
 %a_2}{2(k\cdot p')}~,
 \label{B2}
\end{eqnarray}
 Here, $u_p$ and $\bar u_{p'}$ are free Dirac spinors depending on
 the momenta $p$ and $p'$;
 and $\varepsilon'$ denotes the polarization four vector
 of the scattered photon.
% the effective
% polarization four-vector~\cite{NikishovRitus,Boca-2009} related to
% the polarization vector of scattered photon $\epsilon'_\mu$ as
% $\varepsilon'_\mu=\epsilon'-k'_\mu(\epsilon\cdot k)/(k\cdot k')$
% with properties $\varepsilon'\cdot k'=\varepsilon'\cdot k=0$ and
% $\varepsilon'\cdot \varepsilon'=-1$.
%%%%%%%%%%%%%%%%%%%%%%%%%%%%%%%%%%%%%%%%%%%%%%%%%%%%%%%%%%%%%%%%%%%%%%%%%%%%%%%%%
Utilizing the prescription of Ref.~\cite{TKTH-2013} one can
express the coefficients $C^{(i)}(l)$ through
\begin{eqnarray}
 C^{(0)}(l)&=&\widetilde Y_l(z){\rm e}^{i l\phi_0},\,\,\,\,
 C^{(1)}(l) = X_l(z)\,{\rm e}^{i l \phi_0}~,\nonumber\\
 C^{(2)}(l)&=&\frac{1}{2}\left( Y_{l+1}(z){\rm e}^{i(l+1)\phi_0}
 + Y_{l-1}(z){\rm e}^{i(l-1)\phi_0}\right)~,\nonumber\\
 C^{(3)}(l)&=&\frac{1}{2i}\left( Y_{l+1}(z){\rm e}^{i(l+1)\phi_0}
 - Y_{l-1}(z){\rm e}^{i(l-1)\phi_0}\right)
\label{III25}
\end{eqnarray}
with $\widetilde Y_l(z) = \frac{z}{2l} \left(Y_{l+1}(z) +
Y_{l-1}(z)\right) - \xi^2\frac{u}{u_l}\,X_l(z)$, where
the functions $Y_l(z)$ and $X_l(z)$ are defined by
\begin{eqnarray}
 Y_l(z)&=&\frac{1}{2\pi} \int\limits_{-\infty}^{\infty}\,
 d\psi\,{f}(\psi + \phi_0)
 \,{\rm e}^{il\psi-iz
 {\cal P}(\psi+\phi_0)} ~,\nonumber\\
 X_l(z)&=&\frac{1}{2\pi} \int\limits_{-\infty}^{\infty}\,
 d\psi\,{f^2}(\psi + \phi_0) \,{\rm e}^{il\psi-iz
 {\cal P}(\psi+\phi_0)}~,
 \nonumber\\
 {\cal P(\phi)} &=& z\int_{-\infty}^{\phi}\,d\phi'\,
 \cos(\phi'-\phi_0)f(\phi')
 \nonumber\\
 &-&\xi^2\frac{u}{u_0}
 \int_{-\infty}^\phi\,d\phi'\,f^2(\phi')~.
 \label{III24}
\end{eqnarray}
 The phase $\phi_0$ is equal to the azimuthal angle of the
 direction of flight of the outgoing electron, $\phi_0 = \phi_{e'}$,
 and is related to
 the azimuthal angle of the photon momentum as
 $\phi_{\gamma'}=\phi_0 + \pi$.
 For the dynamical variables in Eqs.~(\ref{III25}) and (\ref{III24})
 we use the standard notation:
 $z=2l\xi\sqrt{\frac{u}{u_l}\left(1-\frac{u}{u_l}\right)}$ with
 $u\equiv(k'\cdot k)/(k\cdot p')$, $u_l=l\,u_0$ and
 $u_0=2{k\cdot p}/m^2$.

 This representation of functions $C^{(i)}(l)$ allows to define a
 partial differential cross section
 \begin{eqnarray}
 \frac{d\sigma(l)}{d\omega'\,d\phi_{e'}}
 =\frac{2\alpha^2}{N_0\,\xi^2\,(s-m^2)\,|p - l\omega|}\,w(l)
  \label{S1}
 \end{eqnarray}
with
\begin{eqnarray}
 w(l)&=&
 -2 \widetilde Y^2_l(z)+\xi^2(1 +\frac{u^2}{2(1+u)})
 \nonumber\\
 &\times& \left(Y^2_{l-1}(z)+ Y^2_{l+1}(z)
 -2\widetilde Y_l(z)X^*_l(z)\right)~.
 \label{S2}
\end{eqnarray}
 Equation (\ref{S2}) resembles the corresponding expression for the partial
 probability of photon emission in the case
 of IPA~\cite{LL4} with the substitutions $l\to n = 1, 2, \cdots$ and
 ${\widetilde Y^2_l(z)},\,Y_l^2(z),\,\widetilde Y_l(z)X^*_l(z)\to J_n^2(z')$,
 namely
 \begin{eqnarray}
 w_n&=&
 -2 J^2_n(z')+\xi^2(1 +\frac{u^2}{2(1+u)})
 \nonumber\\
 &\times& \left(J^2_{n-1}(z')+ J^2_{n+1}(z')
 -2 J_n^2(z')\right)~,
 \label{S2INF}
\end{eqnarray}
 where $J_n(z')$ denotes Bessel functions with
  $z'=\frac{2n\xi}{\sqrt{1+\xi^2}}\sqrt{\frac{u}{u_n}\left(1-\frac{u}{u_n}\right)}$
  and  $u_n=\frac{2n(k\cdot p)}{m^2(1+\xi^2)}$.
 Similarly to IPA, the phase $\phi_0$ can be determined
 through invariants $\alpha_{1,2}$ as $\cos\phi_0=\alpha_1/z$,
 $\sin\phi_0=\alpha_2/z$ with $\alpha_{1,2}=e\left(a_{1,2}\cdot
 p/k\cdot p-a_{1,2}\cdot p'/k\cdot p'\right)$.
%%%%%%%%%%%%%%%%%%%%%%%%%%%%%%%%%%%%%%%%%%%%%%%%%%%%%%%%%%%%%%%%%%%%%%%%%%%

 The dimensionless field intensity $\xi^2$ can be determined
 through the average value of the manifestly covariant variable
 $\eta=T^{\mu\nu}p_\mu p_\nu/(p\cdot k)^2$~\cite{HeinzlIlderton}
 (cf.\ also~\cite{NikishovRitus}),
 where $T^{\mu\nu}$ is the e.m.\ stress-energy tensor
 $T^{\mu\nu}=g_{\alpha\beta} F^{\mu\alpha} F^{\beta\nu}
 + \frac14 g^{\mu\nu} F_{\alpha\beta}F^{\alpha\beta}$ and
 $F_{\mu\nu}=\partial_\mu A_\nu-\partial_\nu A_\mu$ is
 e.m.\ field strength tensor. In the charge's rest frame
 $\eta= T^{00}/\omega^2$, where the
 stress-energy tensor $T^{00}$ is equal to the
energy density of the  e.m.\ field or to
 the pulse intensity $I_L$. %=({\mathbf E^2}+{\mathbf H}^2)/2$.
 In IPA the quantity $\xi^2$ is determined by
 \begin{eqnarray}
 \xi^2=\frac{e^2}{m^2}\frac{1}{\tau_{IPA}}
\int\limits_{-\tau_{IPA}/2}^{\tau_{IPA}/2} dt\,\,
 \eta
 =\frac{e^2}{m^2\omega^2}\frac{1}{2\pi}\int\limits_{-\pi}^{\pi}
 d\phi\,\, I_L %=\frac{e^2a^2}{m^2}~,
 \label{S22}
 \end{eqnarray}
with the above quoted value $\frac{e^2a^2}{m^2}$,
 where the averaging interval is set equal to the duration of one cycle,
 $\tau_{IPA}=2\pi/\omega$. %in an infinite pulse.
The generalization to
 a finite pulse may be done in a straightforward manner:
 \begin{eqnarray}
 \xi^2_{FPA}=\frac{e^2}{m^2}\frac{1}{\tau_{FPA}}\int\limits_{-\infty}^{\infty} dt
 \,\, \eta
 =\frac{e^2}{m^2\omega^2} \frac{1}{2\pi N} \int\limits_{-\infty}^{\infty} d\phi\,\,
 I_L~.
 \label{S23}
 \end{eqnarray}

 Now, the interval $\tau_{FPA}$ is determined by the
 number $N$ of oscillations in a pulse as $2\pi N/\omega$.
 That is, the quantity $\xi^2$, which is included in the expressions
 for the basic functions~(\ref{III24}),
 cross section~(\ref{S1}) and probability~(\ref{S2}),
can be expressed through the averaged value of the intensity
 of a finite laser pulse
 \begin{eqnarray}
 \xi^2=\xi^2_{FPA} \frac{N}{N_0}~,
 \label{S24}
 \end{eqnarray}
or
\begin{eqnarray}
  \xi^2=\frac{N}{N_0}\frac{e^2}{\omega^2m^2}\langle I_L  \rangle
  \simeq \frac{N}{N_0} \frac{5.6\cdot 10^{-19}}{\omega^2[eV^2]}
  \langle I_L \rangle \left[\frac{\rm W}{\rm  cm^2}\right]~,
  \label{S244}
\end{eqnarray}
 where
 $N\, \langle I_L\rangle=(\omega/2\pi)\int_{-\infty}^{\infty}dt\,I_L$.
 Hence, the normalization
 factor $N_0$ defined as
\begin{eqnarray}
 N_0=\frac{1}{2\pi}\int\limits_{-\infty}^{\infty}
 d\phi\,(f^2(\phi)+ {f'}^2(\phi))
 \label{S25}
\end{eqnarray}
 has the meaning to renormalize the photon flux in case of
 the finite pulse and to determine the cross section in Eq.~(\ref{S1}).
 The factor $N_0$ is described in some detail in Sect.~III.A below.
 In fact, for the considered envelope functions
 $N_0\simeq N$ and, therefore, $\xi^2\simeq \xi^2_{FPA}$.
%%%%%%%%%%%%%%%%%%%%%%%%%%%%%%%%%%%%%%%%%%%%%%%%%%%%%%%%%%%%%%%%%%%%%

 The frequency $\omega'$ of the emitted photon is related to the
 auxiliary variable $l$  and the polar angle $\theta'$
 of the direction of the momentum $\vec k'$ via
\begin{eqnarray}
 \omega'=\frac{l\,\omega (E+|\vec p|)}{E + |\vec p| \cos\theta'
 +l \omega(1-\cos\theta') }
 \label{S3}
\end{eqnarray}
 and increases with $l$ at fixed $\theta'$ since $\omega'$ is
 a function of $l$ at fixed $\theta'$. For convenience,
 we also present a similar expression for IPA,
 where the fermions are dressed and the integer
 quantity $n$, together with the field intensity $\xi^2$, appear:
 \begin{eqnarray}
 \omega'=\frac{n\,\omega (E+|\vec p|)}{E + |\vec p| \cos\theta'
 +\omega(n + \frac{m^2\xi^2}{2(k\cdot p)} )(1-\cos\theta')}~.
 \label{S3_}
\end{eqnarray}
% For reasoning of $l$ as an internal variable cf.\cite{DSeipt-2014}

The differential cross section of the one-photon production is
eventually
\begin{eqnarray}
 \frac{d\sigma}{d\omega'}=\int\limits_{\zeta}
 dl\,\int\limits_{0}^{2\pi}d\phi_{e'}
 \frac{d\sigma (l)}{d\omega' d \phi_{e'}}
 \delta \left(l-l(\omega') \right)~.
 \label{S33}
\end{eqnarray}
 The lower integration limit $\zeta >0$
 is defined by kinematics, i.e.\ by the minimum value of
 the considered $\omega'$, in accordance with Eq.~(\ref{S3}). In
 the IPA case, the variable $n = 1, 2,\, \cdots$ refers to
 the contribution of the individual harmonics
 ($n = 1$ with $\xi^2\ll 1$ recovers the Klein-Nishina cross
 section, cf.~\cite{Ritus-79}). The value $n\omega$ is
related to the energy
 of the background field involved in Compton scattering.
 Obviously, this value is a multiple of $\omega$.
 In FPA, the internal quantity $l$ is a continuous
 variable, implying a continuous distribution of the differential
 cross section over the $\omega' - \theta'$ plane.
 The quantity $l\omega$ can be considered as energy
 of the laser beam involved in the Compton process,
 which is not a multiple $\omega$.
 Mindful of this fact, without loss of generality, we
 denote the processes with $l>1$ as a multi-photon
 generalized Compton scattering, remembering
 that $l$ is a continuous quantity.
%(for more details, cf.~\cite{DSeipt-2014}).

%%%%%%%%%%%%%%%%%%%%%%%%%%%%%%%%%%%%%%%%%%%%%%%%%%%%%%%%%%%%%%%%%
 The multi-photon effects become most clearly evident in the
 partially energy-integrated cross section
 \begin{eqnarray}
 {\tilde\sigma_{}(\omega')} = \int\limits_{\omega'}^{\infty}
 d\bar\omega' \frac{d\sigma (\bar\omega')}{d\bar\omega'}
 =\int\limits_{l'}^{\infty} dl
 \frac{d\sigma(l)}{dl}~,
 \label{S6}
\end{eqnarray}
 where
 $d\sigma(l) / dl =
 ( d\sigma(\omega') / d\omega')
 (d\omega'(l) / dl)$,
 and the minimum value of $l'$ is
 \begin{eqnarray}
 l'=\frac{\omega'}{\omega}\,
 \frac{E+|\vec p|\cos\theta'}{E+|\vec p|-\omega'(1-\cos\theta')}~.
 \label{S66}
 \end{eqnarray}
 The cross section (\ref{S6}) has the meaning of a cumulative distribution.
 In this case, the subthreshold, multi-photon events correspond to
 frequencies $\omega'$ of the outgoing photon which
 exceed the corresponding
 threshold value $\omega_1'=\omega'(l=1)$ (cf. Eq.~(\ref{S3})).

%%%%%%%%%%%%%%%%%%%%%%%%%%%%%%%%%%%%%%%%%%%%%%%%%%%%%%%%%%%%%%%%%%%%%%%%%%%%%

\section{numerical results}
%\subsection{The e.m. potential $\mathbf A$ and field strength $\mathbf E$}
\subsection{The envelope shapes and the e.m. field structure}

 Some aspects of the one- and two-parameter envelope
 functions~(\ref{E1}) have been considered
 in~\cite{TKTH-2013}. Here we extend this analysis.
 The parameter $\Delta$ characterizes the
 pulse duration $2\Delta$ with $\Delta=\pi N$,
 where $N$ has a meaning
 of a "number of oscillations" in the pulse.
 Certainly, such a definition is rather conditional
 and is especially meaningful for the flat-top envelope
 with small values of $b/\Delta$. In the case of the
 hs envelope shape, the number of oscillations
 with small amplitudes may exceed $N$. Nevertheless,
 for convenience we call $N$ "number of oscillations
 in a pulse" for given $f(\phi)$, relying on its relation with
 the shape parameter $\Delta$.
 The parameter $b$ in
 the two-parameter sF shape has the meaning of the "thickness"
 or ramping time of the pulse shape. It was shown that
 the properties of the two-parameter sF shape for
 large values of $b/\Delta\simeq0.3\dots0.5$ are close to that of
the one-parameter hs shape. Therefore, as mentioned above,
 in order to stress the difference between one- and
 two-parameter (flat-top) envelopes we focus our
 consideration on the choice of $b/\Delta=0.15$
troughout our paper.
%%%%%%%%%%%%%%%%%%%%%%%%%%%%%%%%%%%%%%%%%%%%%%%%%%%%%%%%%%%%%%%%%%%%

\begin{figure}[h!]
 \includegraphics[width=0.45\columnwidth]{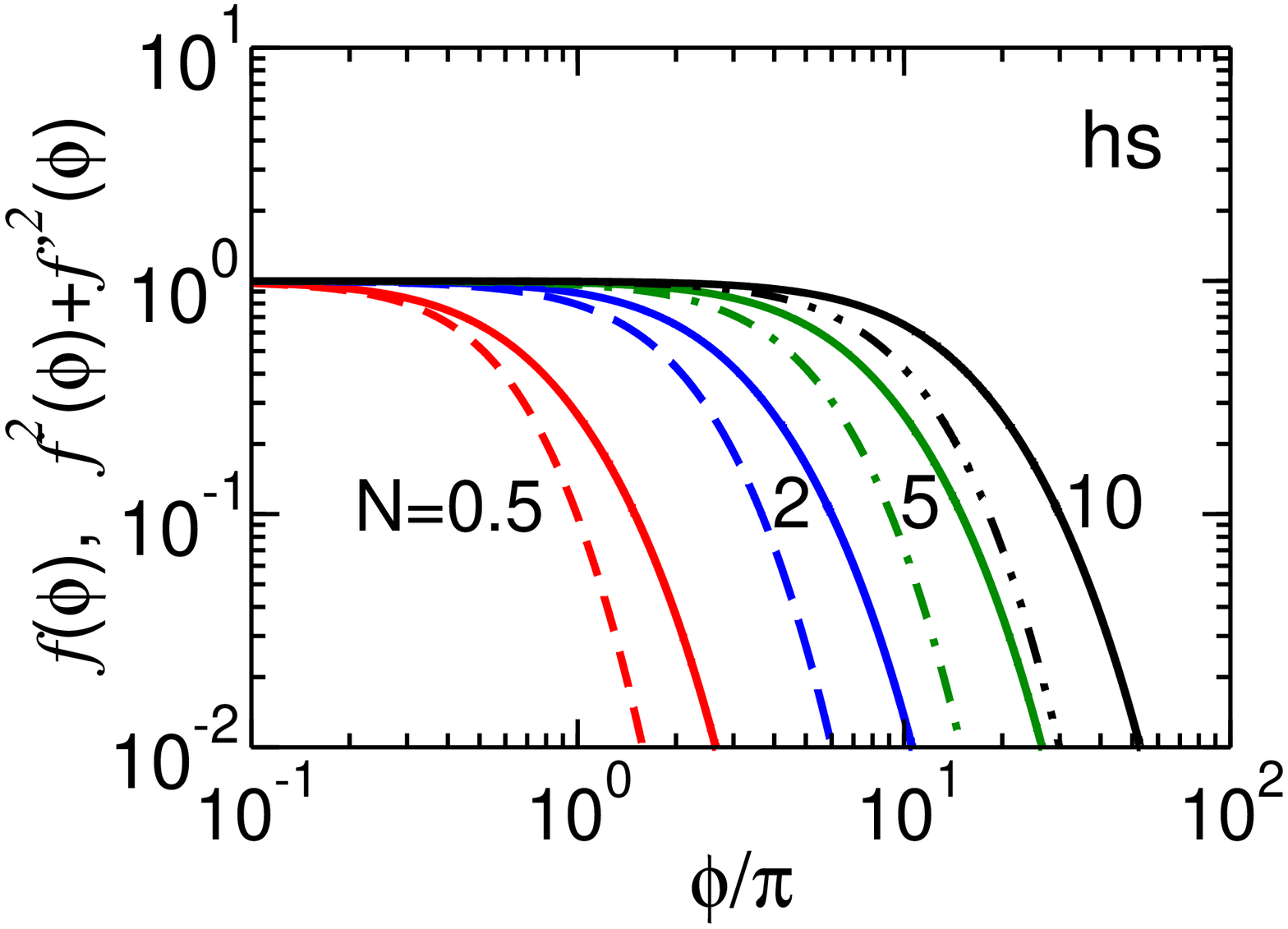}\qquad
 \includegraphics[width=0.45\columnwidth]{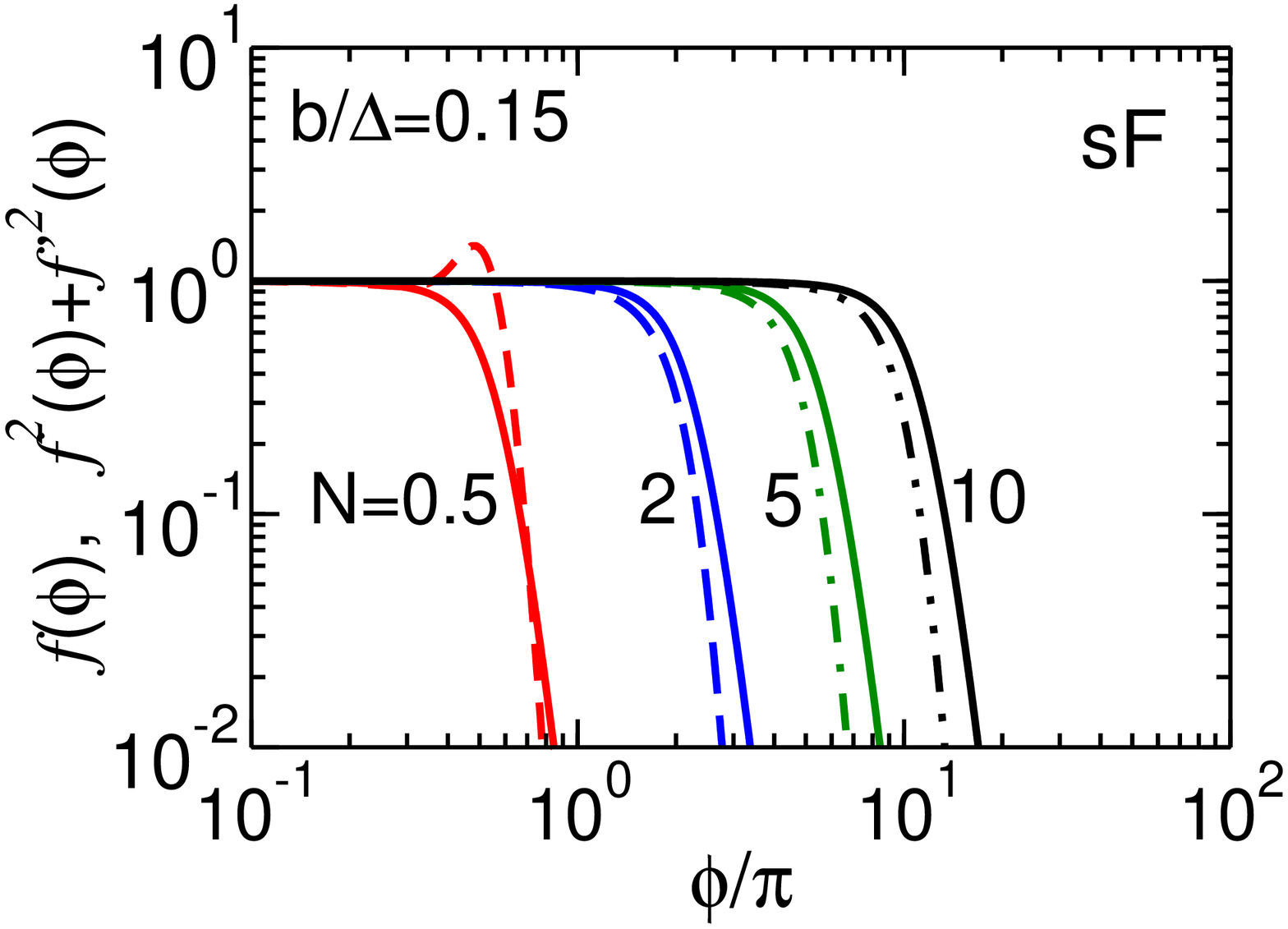}
 \caption{\small(Color online)
 The envelope functions $f(\phi)$
 and the integrand  $f^2(\phi)+{f'}^2(\phi)$ in Eq.~(\ref{S25})
 as the functions of invariant phase $\phi=kx$.
 The thick solid curves
 labeled by $N$ are for $f(\phi)$.
 The dashed, long-dashed, dot-dashed and dot-dot-dashed curves are for
 $f^2(\phi)+{f'}^2(\phi)$ with $N=0.5$, 2, 5 and 10, respectively.
 Left and right panels are for hs and  sF envelope  shapes,
 respectively. % For the later one $b/\Delta=0.15$.
 \label{Fig:01} }
 \end{figure}

 The envelope shape $f(\phi)$ and the integrand  $f^2(\phi)+ {f'}^2(\phi)$
 in Eq.~(\ref{S25}) as functions of the invariant phase for hs and sF shapes
 are  shown in  Fig.~\ref{Fig:01} in left and right
 panels, respectively.
 The numbers in the plot indicate the number $N$ of
 oscillations in a pulse. The thick solid curves
 labeled by $N$ are for $f(\phi)$.
 The dashed, long-dashed, dot-dashed and dot-dot-dashed curves are for
 $f^2(\phi)+{f'}^2(\phi)$ with $N=0.5$, 2, 5 and 10, respectively.
% In case of sF shape the ratio $b/\Delta=0.15$,

 For the smooth  hs shape the integrand is also a smooth
 function (cf.\ Fig.~\ref{Fig:01}, left panel).
 For the flat-top sF envelope shape and  $N \ge 2$
 both, $f(\phi)$ and the integrand $f^2(\phi)+{f'}^2(\phi)$
(the latter one being proportional to the intensity in the course of the pulse),
are smooth functions of the invariant phase
 which more compact as compared with the hs shape with
 the same value of $N$. At $N=0.5$ and $\phi\sim\Delta$ the integrand
(see dashed red curve in the right panel) displays some
 %"bump-like" behavior
overshoot resulting locally in the height
 $h=1/4 + (\frac{\Delta}{b}/4\Delta)^2\simeq 1.37$.
 Increasing $\Delta$ (or $b/\Delta$) leads
 to a vanishing of this overshoot.

 For the hs envelope, the normalization factor in Eq.~(\ref{S25}) has the form
 \begin{eqnarray}
 N^{\rm hs}_0=\frac{\Delta}{\pi}\left(1+  \frac{1}{3\Delta^2}
 \right)~,
 \label{EE5}
 \end{eqnarray}
while for the sF shape one has
\begin{eqnarray}
N^{\rm sF}_0=\frac{\Delta}{\pi}\left( F_1\left(t \right)
 +  F_2\left(  t \right) \frac{b}{\Delta}\right),\,\,\,
 t =\frac{  1 + \cosh\frac{\Delta} {b} }  {\sinh\frac{\Delta} {b}} ,
\label{EE6}
\end{eqnarray}
where
\begin{eqnarray}
 F_1(t)&=&\frac{(t^2+1)(-t^4 +10 t^2 -1)}{16t}~,\nonumber\\
 F_2(t)&=&\frac {3t^{10}-  35t^8+  90t^6-  90t^4+
 35t^2-3}{24(t^2-1)^3}~.
 \label{EE7}
\end{eqnarray}
In the limit  $\frac{b}{\Delta}\to 0 $,
\begin{eqnarray}
 N_0^{\rm sF} = \frac{\Delta}{\pi} +
 {\cal O}\left(\exp[-\frac{\Delta}{b}] \right) \simeq  \frac{\Delta}{\pi} .
 \label{EE8}
\end{eqnarray}

\begin{figure}[h!]
\includegraphics[width=0.55\columnwidth]{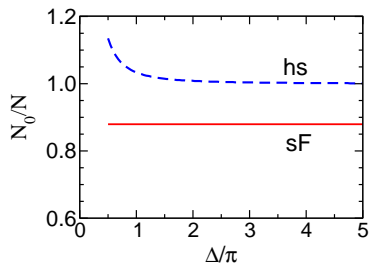}
 \caption{\small(Color online)
 The scaled normalization factor $N_0/N$
 as a function of the number of oscillations in the pulse,
 $N={\Delta}/{\pi}$, for hs and sF shapes,
 shown by the dashed blue and
 solid red curves, respectively.
% In later case $b/\Delta=0.15$
 \label{Fig:02} }
 \end{figure}

 The normalization factor $N_0$
 scaled by $N=\Delta/\pi$ as a function of $N$
 for hs and sF shapes is exhibited in Fig.~\ref{Fig:02},
 shown by the dashed blue and
 solid red curves, respectively.

 For the hs shape, $N_0\simeq N$ at $N\ge1$
 and slightly increases for the sub-cycle envelopes with $N<1$
 (cf. Eq.~(\ref{EE5})).
 In the case of a flat-top envelope, the ratio $N_0/N$ is independent of
 $\Delta$, according to Eq.~(\ref{EE7}).
 The contribution of ${f'}^2$
 in~(\ref{S25}) is weak and varies
 from 0.2\%
 to 3.8\%  for $b/\Delta=0.01$ and 0.2, respectively.
 In the limit $b/\Delta\to 0$ it vanishes and
 $N_0\to N$ and, therefore, the overshoot in the integrand does
 not affect the integral in Eq.~(\ref{S25}).
 But taking into account that very small values of $b/\Delta$
 seems to be not realistic, we restrict our actual calculations
 to the finite value $b/\Delta=0.15$,
 where the overshoot in $f^2(\phi)+ {f'}^2(\phi)$ is minor.

%%%%%%%%%%%%%%%%%%%%%%%%%%%%%%%%%%%%%%%%%%%%%%%%%%%%%%%%%%%%%%%%%%%%%
 For the sake of completeness,
 we present also the behavior of
 e.m.\ potential $\vec A$
 and the electric field strength
 ${\vec E}= - \partial {\vec A} / \partial t$,
 where ${\vec A}$ is
 given by Eqs.~(\ref{III1}) and (\ref{E1}) as functions of
 the invariant phase $\phi$.
 The e.m. potential and strength
 for the one- and two-parameter envelope functions read
\begin{eqnarray}
 {A_x}&=&a\,f(\phi)\cos\phi~,\qquad
 {A_y} = a\,f(\phi)\sin\phi~,
 \label{EEA}\\
 {E_x}&=&\omega\, {A_x}
 \left[
 -(\ln f(\phi))' + \tan\phi
 \right]~,\label{EEx}\\
 {E_y}&=&\omega\, {A_y}
 \left[
 -(\ln f(\phi))' - \cot\phi
 \right]~,
 \label{EEy}
\end{eqnarray}
with
\begin{eqnarray}
 -(\ln f(\phi))'=
 \left\{
\begin{array}{ll}
 \frac{1}{\Delta}\tanh\frac{\phi}{\Delta},&{\small\rm hs},\\
 \frac{1}{b}\frac{\sinh{\frac{\phi}{b}}}{\cosh\frac{\Delta}{b}+\cosh\frac{\phi}{b}},&
 {\small\rm sF}.
\end{array}
 \right.
 \label{EE2}
\end{eqnarray}

\begin{figure}[h!]
\includegraphics[width=0.45\columnwidth]{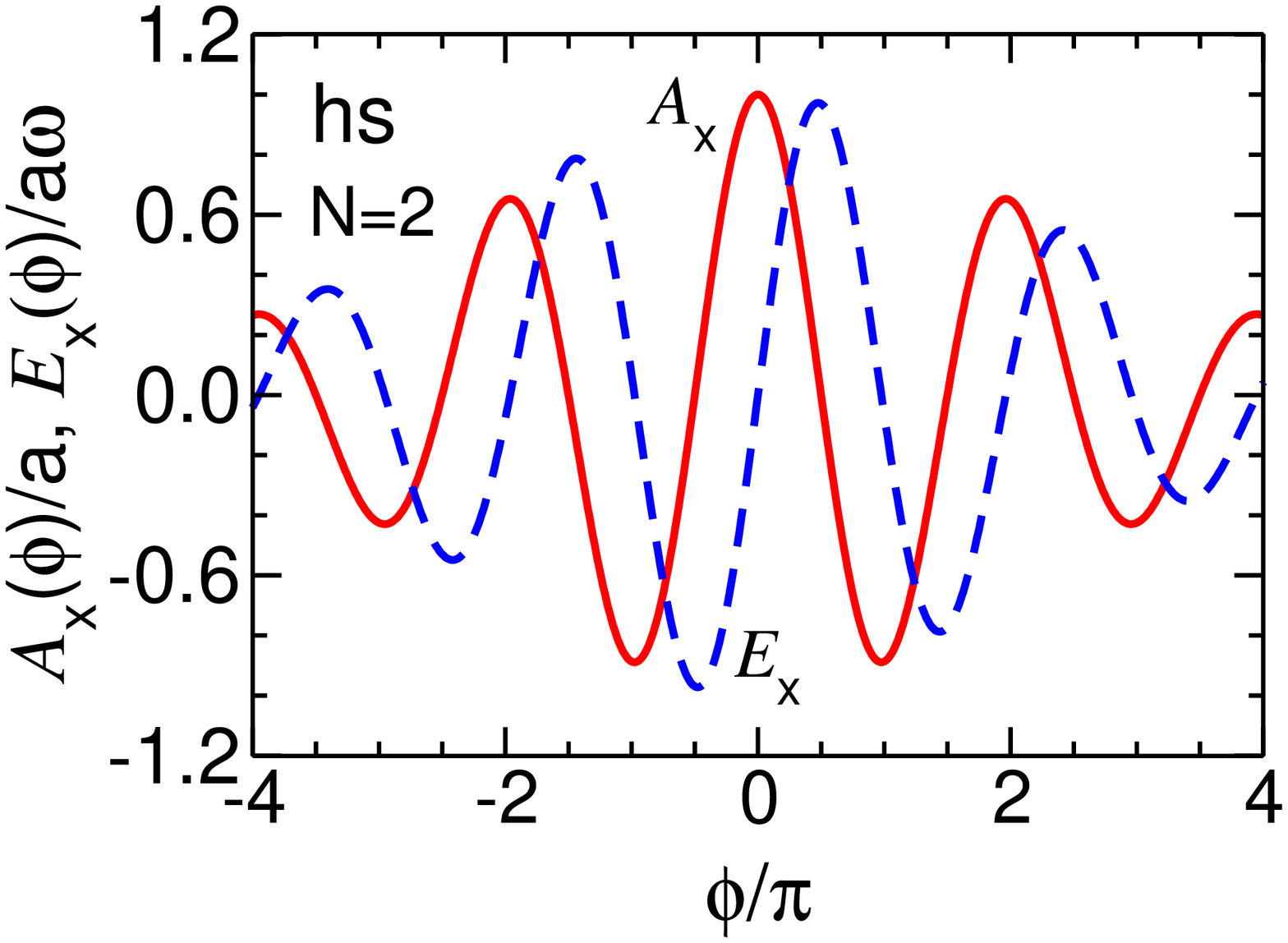}\qquad
\includegraphics[width=0.45\columnwidth]{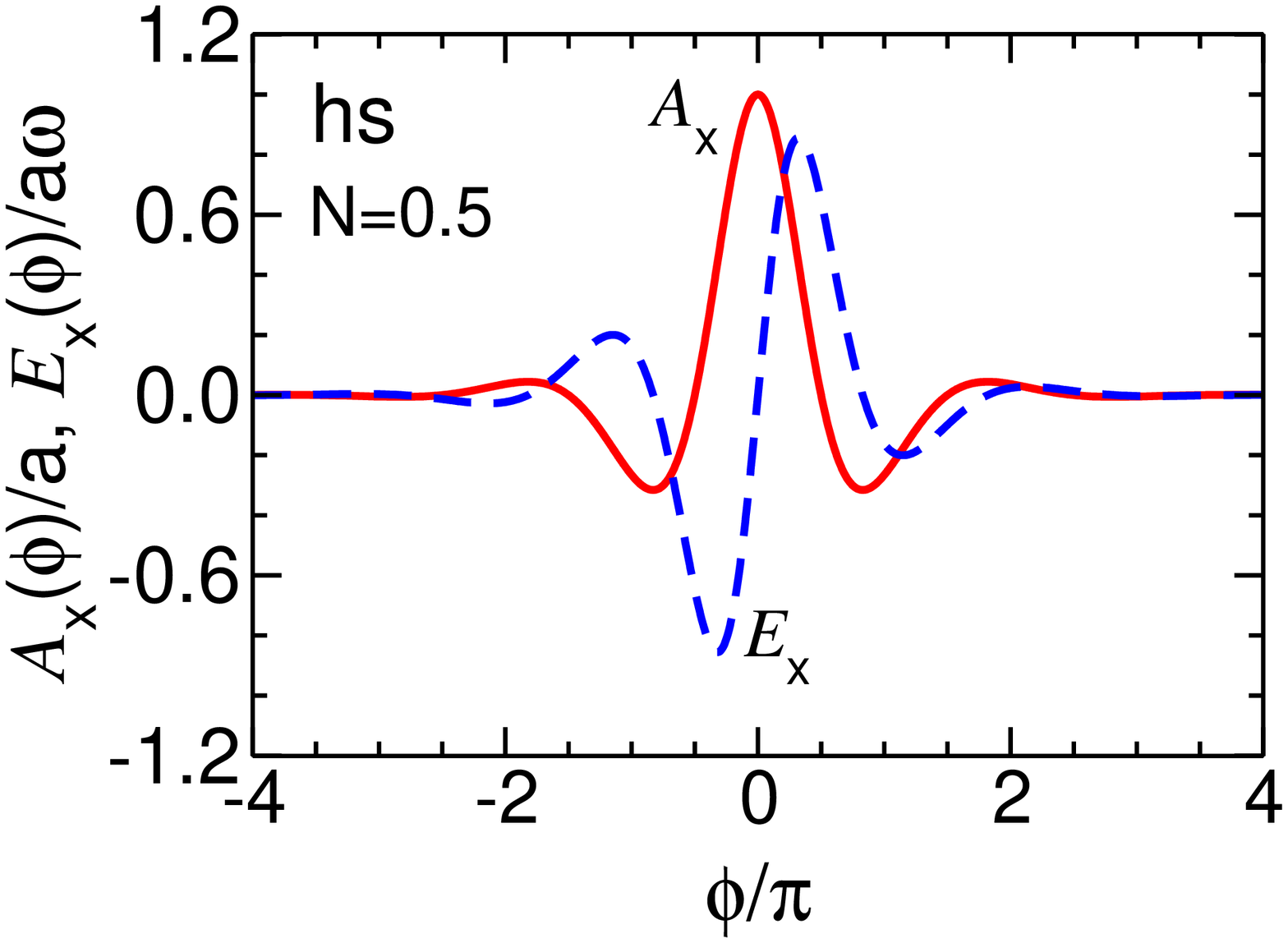}\\
\includegraphics[width=0.45\columnwidth]{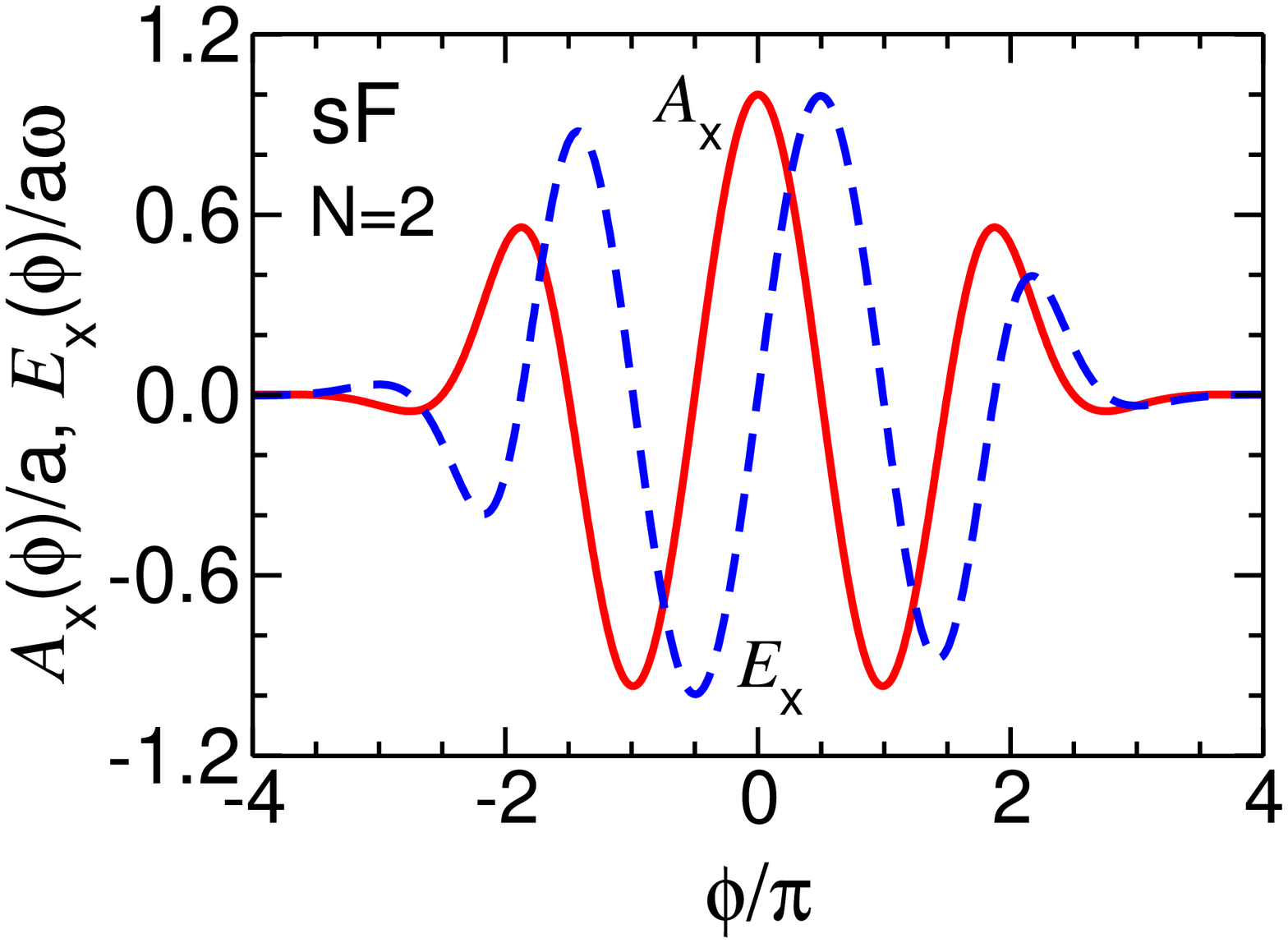}\qquad
\includegraphics[width=0.45\columnwidth]{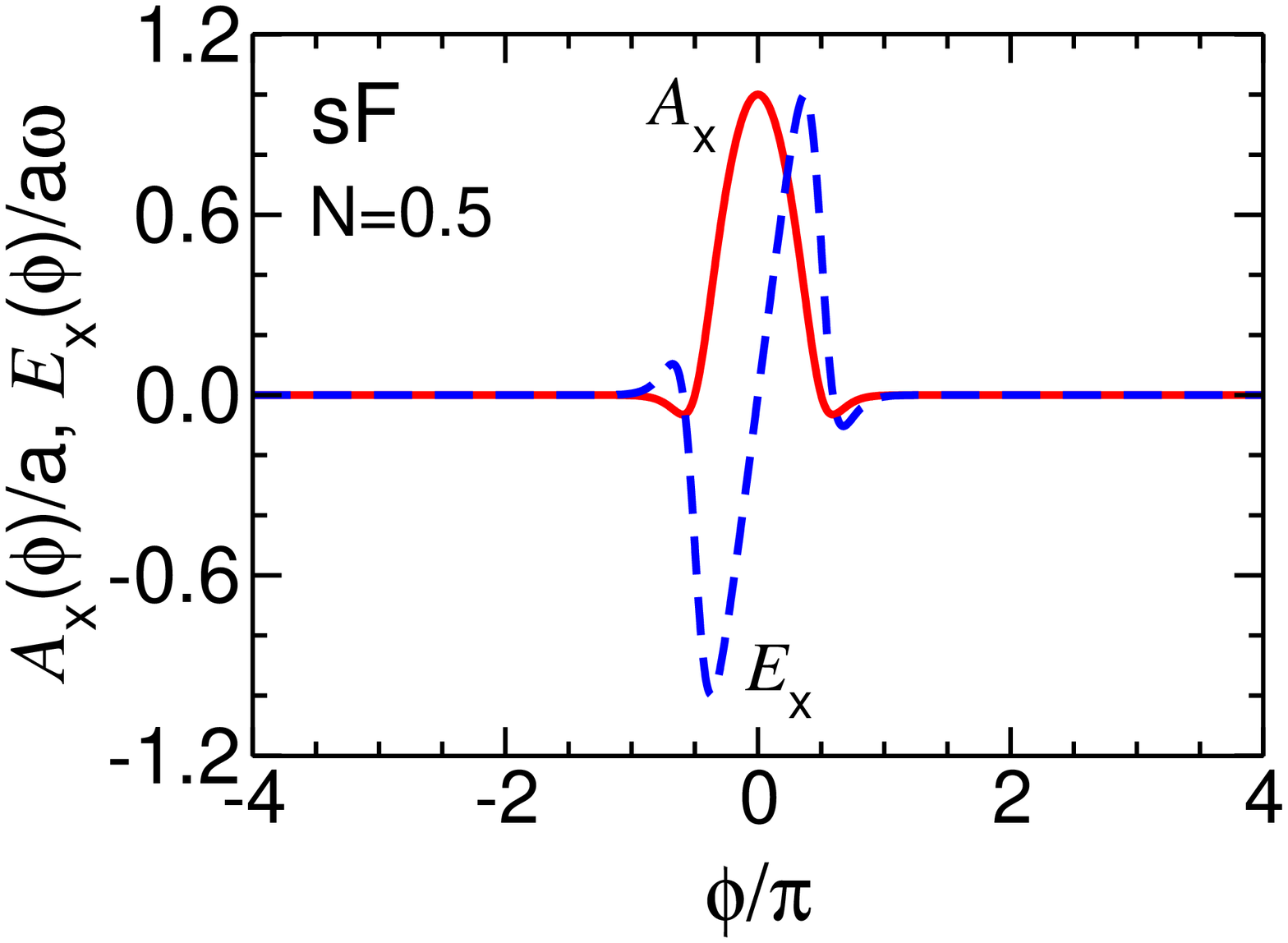}\\
\includegraphics[width=0.45\columnwidth]{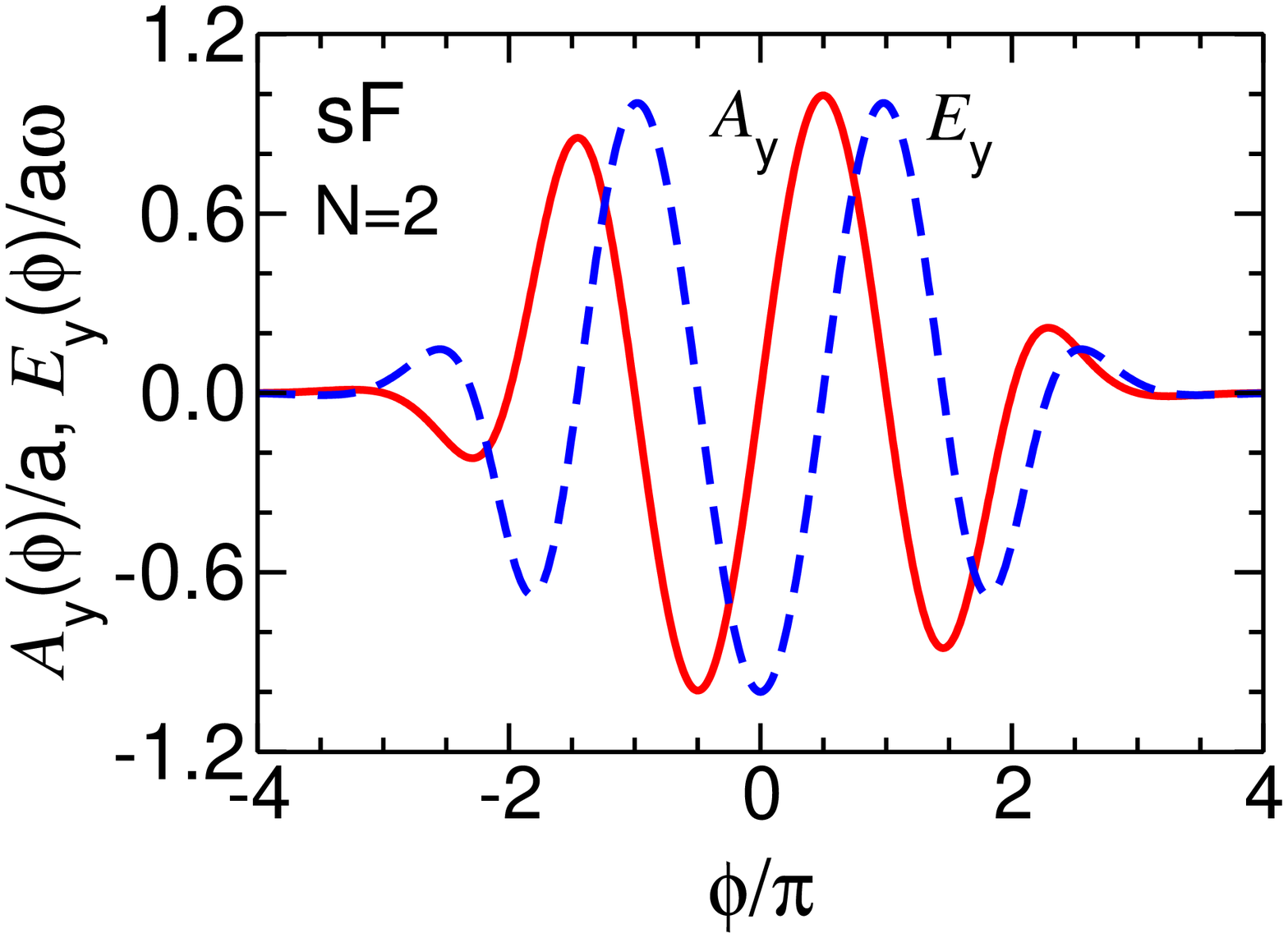}\qquad
\includegraphics[width=0.45\columnwidth]{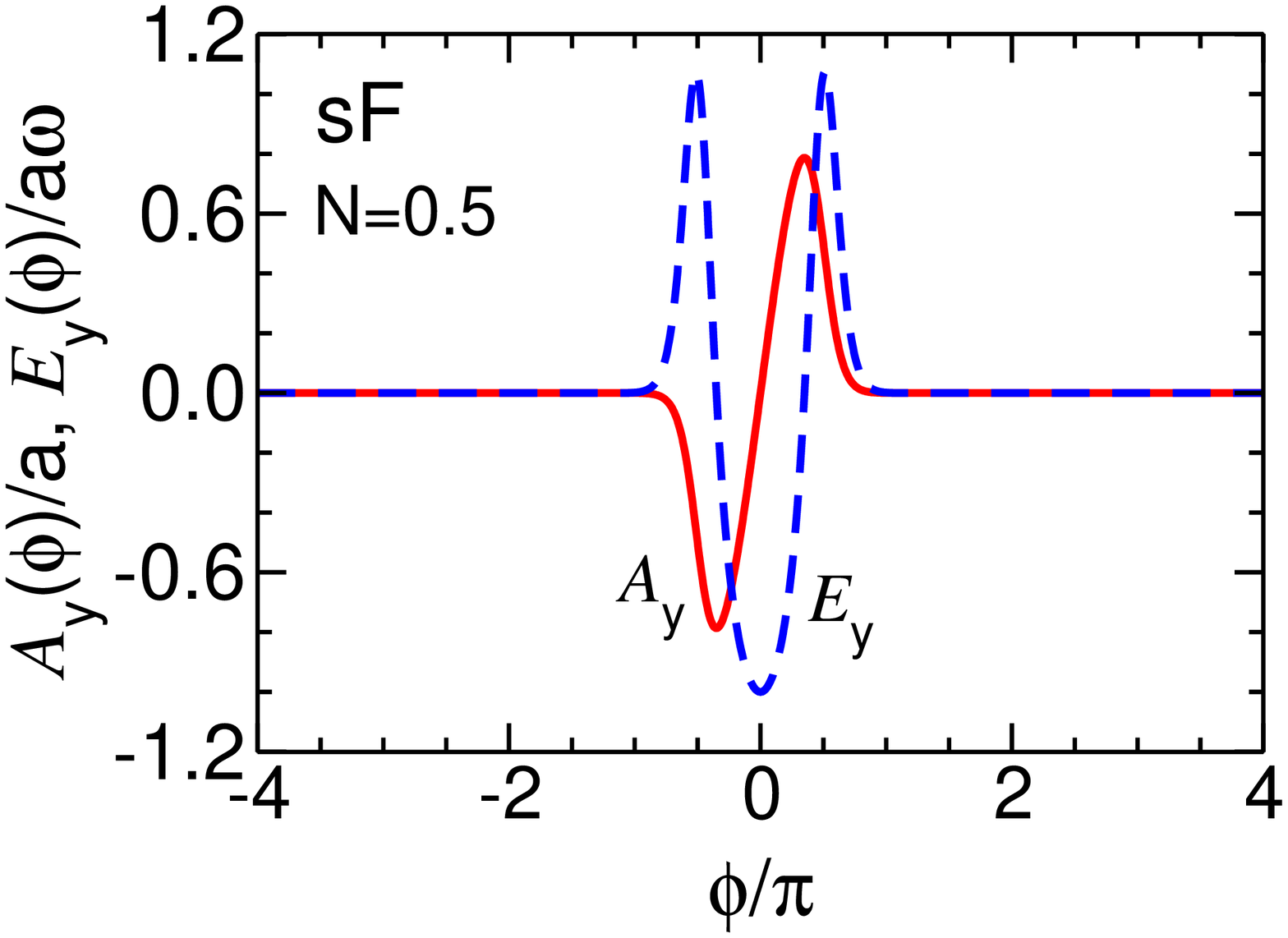}
 \caption{\small(Color online)
 The e.m.\ potentials $A/a$ (solid red curves) and field
 strengths $E/a\omega$ (dashed blue curves) as functions of the invariant
 phase $\phi$. The upper and middle panels correspond to the
 hyperbolic secant (hs) and symmetrized Fermi (sF) shapes,
 respectively, for the $x$ components.
 The lower panels correspond to the $y$ components
 for the sF envelope shape.
 %In all cases for sF envelope, $b/\Delta=0.15$.
 The left and right panels are for pulses with $N=2$ and $N=0.5$,
 respectively.
 \label{Fig:03} }
\end{figure}

 The scaled potentials $A_x/a$ and the scaled
 strengths $E_x/a\omega$ as functions of the invariant
 phase are exhibited by solid red and dashed blue curves,
 respectively, in upper and middle panels in Fig.~\ref{Fig:03}
 for the hs and sF shapes.
 The left and right panels correspond to the pulses with
 $N=2$ and 0.5, respectively.
 The result for the hs shape with $N=0.5$ is close qualitatively  to that
 of~\cite{Mackenroth-2011}.
 One can see that the duration of the pulse increases with
 increasing number of oscillations. The
 flat-top sF shape is more compact compared to the hs shape with
 the same value of the pulse "scale" parameter $\Delta$.

 The result for $y$ components is exhibited in Fig.~\ref{Fig:03}, lower
 panels, where we restrict ourselves to the example of the
 flat-top sF envelope shape. For short pulses with $N>2$, the
 contribution of the first terms in Eqs.~(\ref{EEx}) and (\ref{EEy})
 are relatively small and, therefore, the approximate relations
 $A_y\simeq E_x/\omega$ and $E_y\simeq -\omega\,A_x$ are valid.
 Both $A_y$ and $E_y$ are finite. The same is valid for sF shape
 with $N\ge 2$.
 This is illustrated in Fig.~\ref{Fig:03}, lower panel (left), where the
 result for the sF envelope with $N=2$ is shown.
 The approximate relations are valid also
 for sub-cycle pulse with $N=0.5$
 and the one-parameter hs shape. In the case of the flat-top envelope
 for $N=0.5$, the above approximate
 relations are valid for $A_y$ and for the central part of $E_y$
 (cf. Fig.~\ref{Fig:03} lower panel (right)).
 In the border area with $\phi\approx\Delta=\frac{\pi}{2}$,
 the strength $E_y$ has finite narrow peaks with
 height $\tilde h=\left(\frac{\Delta}{b}\right)
 \frac{\sin\Delta}{4\Delta} +{\cal O}(\exp(-\Delta/b))
 \simeq 1.06$. The height of these peaks
 decreases with increasing $\Delta$
 at fixed $b/\Delta$ and
 for $N\ge2$ it becomes negligibly small.
 This "pick-like" behavior for the flat-top
 shape can be compared with the
 the popular rectangular pulse~\cite{Boca-2009} where the derivative
 $f'(\phi)=\theta'(\phi-\Delta)=\delta(\phi-\Delta)$ is singular
 at $\phi=\Delta$.
 But such "pick-like" or even singular behavior of $\vec E$ at
 the border does not affect the transition matrix $M(l)$ in Eq.~(\ref{EM}),
 because it is determined  by $A$ and $A^2$
 rather than the e.m.\ strength.
% But to avoid confusion, as we mentioned above,
% we limit our actual calculation by the finite value
% of $b/\Delta=0.15$ where $E_y$ is finite.

% To summarize this part we note that in case
% of the finite laser pulse
% the gauge invariant e.m. potential $A_\mu$
% proportional to the envelope functions $f(\phi)$
% solely determine the transition matrix, while the
% e.m. strength $\mathbf E$ is responsible
% for the gauge invariant determination
% dimensionless e.m. field intensity $\xi^2$
% trough the overall normalization factor $N_0$
% which is related to the average laser intensity.

\subsection{The differential cross sections}

In IPA \cite{Ritus-79,LL4}, the cross section of the multi-photon
Compton scattering increases with $\theta'$ towards $180^o$.
For instance, it peaks at about $170^o$ for the chosen
electron energy of 4 MeV (all quantities are considered in the
laboratory frame) and rapidly drops to zero when $\theta'$
approaches $180^o$ for the harmonics $n > 1$ yielding thus the
blind spot for back-scattering. Therefore, in our subsequent
analysis we choose the near-backward photon production at $\theta'
= 170^o$ and an optical laser with $\omega=1.55$~eV. Defining
one-photon events by $n = 1$, this kinematics leads via
Eq.~(\ref{S3_}) to $\omega_1'\equiv \omega'(n=1,\xi^2\ll1,\,
\theta' = 170^o) \simeq 0.133$~keV which we refer to as a
threshold value. Accordingly, $\omega' > \omega'_1$ is enabled by
non-linear effects, which in turn may be related loosely to
multi-photon dynamics with $n>1$ in IPA or $l>1$ in FPA where,
 we remind again, the internal
 variable $l$ can not be interpreted strictly as number of laser
 photons involved (cf.~\cite{DSeipt-2014}).
% In addition, short pulses have a finite bandwidth, furthermore
% enhancing the yield at $\omega' > \omega'_1$ due to the higher
% frequencies in the laser-pulse power-spectrum.
 Note that all calculations for IPA
 are performed in a standard way~\cite{Ritus-79,LL4}.
 The energy of the outgoing photon in IPA is calculated using
 Eq.~(\ref{S3_}), where dressing of electrons in the
 background field is taken into account.

\begin{figure}[h!]
\includegraphics[width=0.95\columnwidth]{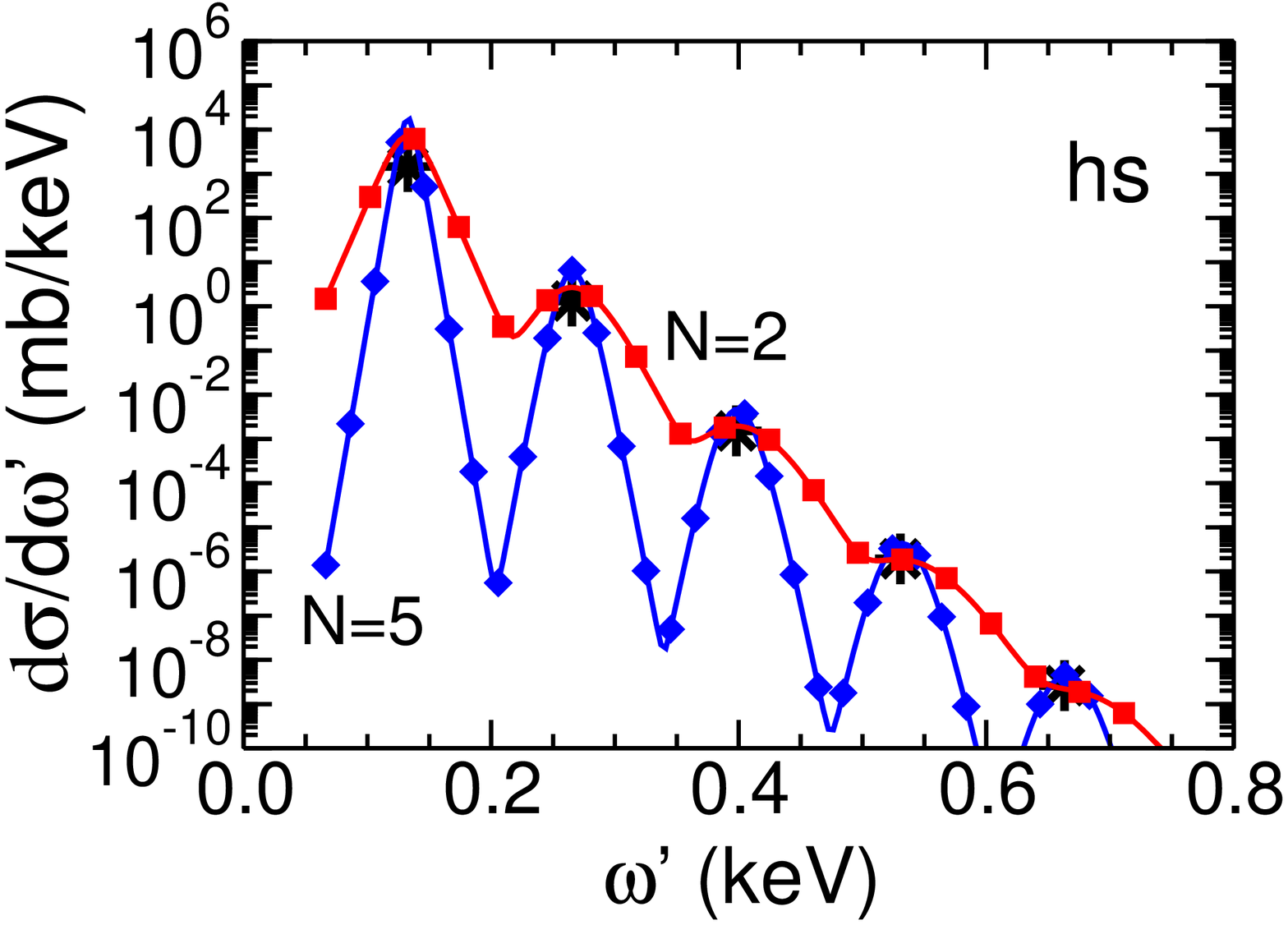}\qquad
\includegraphics[width=0.95\columnwidth]{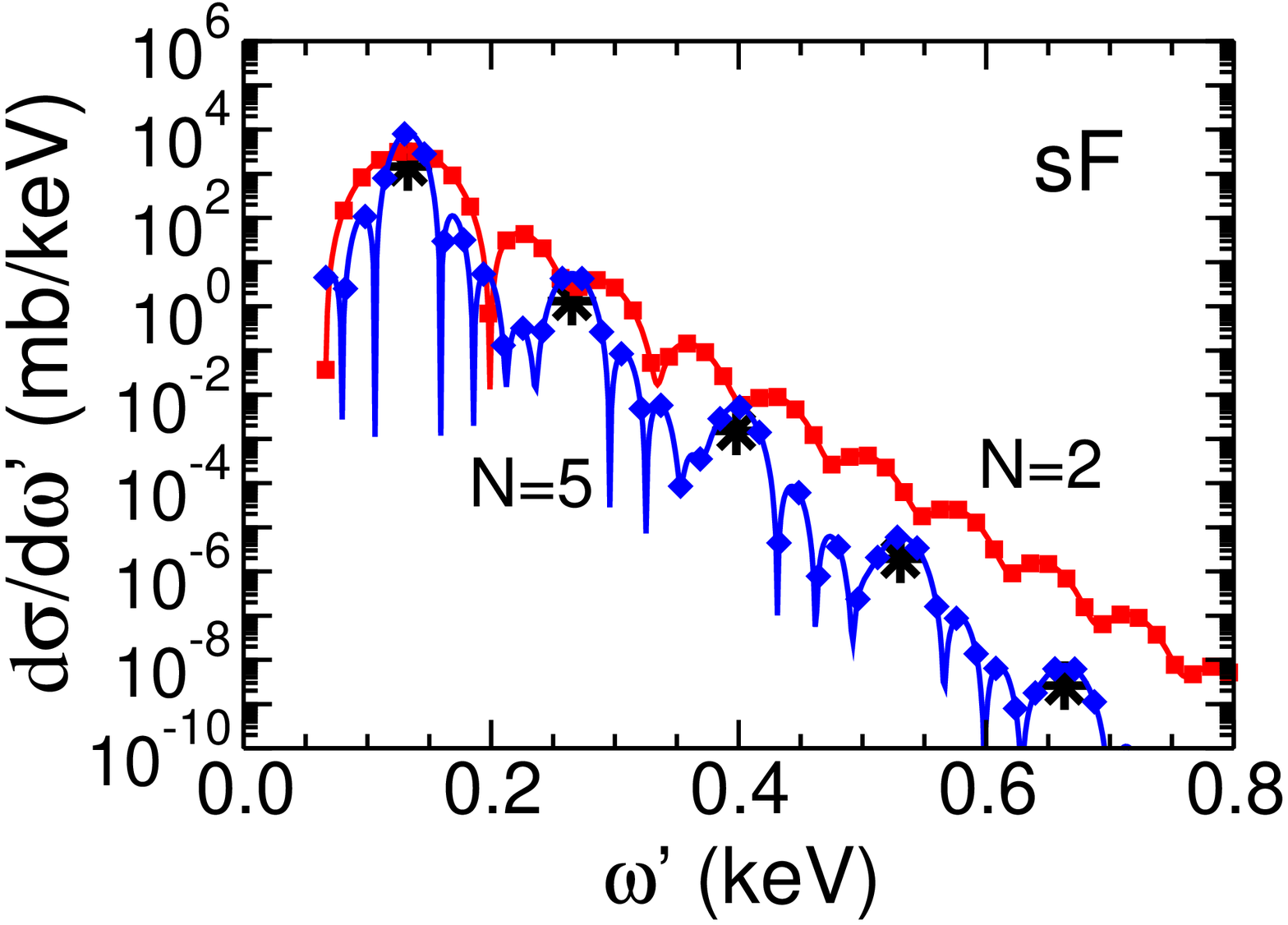}

 \caption{\small(Color online) Differential cross section $d \sigma / d\omega' \,
 \vert_{\theta' = 170^o}$ of Compton scattering for
 $\xi^2=10^{-3}$.
 The red curves (marked by boxes)  and  blue curves
(marked by diamonds) are for $N = 2$ and  5, respectively.
 The black stars depict the IPA results for lowest
 harmonics. Top and bottom panels
 correspond to hyperbolic secant (hs) and symmetrized Fermi (sF)
 shapes of the envelopes, respectively.
%In the latter case, $b/\Delta=0.15$.
 \label{Fig:4} }
\end{figure}

 Let us consider first an example of short pulses
 with moderate intensity, $\xi^2 = 10^{-3}$,
 similar to a recent experiment
 of Compton backscattering~\cite{Jochmann}.
 Results for the hs and sF
 shapes are exhibited in Fig.~\ref{Fig:4}.
% For the two-parameter sF shape, all results
% shown below are for
% $b/\Delta=0.15$ with $\Delta=\pi N$, where the number of
% oscillations $N$ in a pulse is taken as input.
 The red curve (marked by boxes) and the blue curve
 (marked by diamonds)
 correspond to pulses with
 $N=2$ and 5, respectively.
 The black stars depict the IPA results, i.e., the harmonics at fixed
 scattering angle $\theta'$.
 Their positions correspond to integer values of
 $n=1, \,2, \cdots$ in accordance with Eq.~(\ref{S3_}).
 i.e.\ the distribution of scattered photon energies is
 a discrete function of $\omega'$.
 We stress that the cross section at $\omega'>\omega_1'$
 is essentially "sub-threshold", i.e.\ outside the kinematically
 allowed region of the Klein-Nishina process due to multi-photon effects.

 In the FPA case, the energy distribution
 becomes a continues function of $\omega'$. The actual shape
 is determined by both the pulse duration and the envelope form.
 Consider first the case of the hs pulse
 (cf.~Fig.~\ref{Fig:4}, top panel).
 The cross section displays sharp bumps with peak positions
 corresponding to integer values of $l=n$ (as in IPA).
 In the vicinity of the bumps, at $l=n \pm \epsilon$,
 $\epsilon \ll 1$,
 the cross section is rapidly decreasing.
 Such a behavior reflects the properties of the functions
 $Y_l(z)$~\cite{TKTH-2013}
 which behave under such conditions as
 \begin{eqnarray}
 Y_{n+\epsilon}(z)&\simeq& \frac{z^n}{2^nn!}\,{\rm
 e}^{-i\epsilon\phi_0} F^{(n+1)}(\epsilon)~,
 \label{B6}
\end{eqnarray}
where $F^{(n)}(\epsilon)$ is the Fourier transform of the function
$f^n(\phi)$. At $\xi^2 \ll 1$, the contribution of terms $\propto
X_l$ is negligible. The behavior of the cross section in the
vicinity of the first bump is proportional to $F^2_{\rm
hs}(\epsilon)$ with
\begin{eqnarray}
 F_{\rm hs}(x)&=&\frac{\Delta}
 { 2\cosh  {\frac12\pi\Delta  |x|}} ~,
 \label{S4}
 \end{eqnarray}
 or
 $ F_{\rm hs}(x)\simeq \Delta \exp[-\pi\Delta x/2]$.
 Thus,  the cross section becomes steeper
 with increasing pulse duration $\Delta$. This result
 qualitatively agrees with that of Ref.~\cite{Krajewska-2012}.

 In the case of the
 sF shape, the dependence $F_{\rm sF}(\epsilon)$ is more complicated:
\begin{eqnarray}
 F_{\rm sF}(x)&=&\frac{1+{\exp}\left[{-\frac{\Delta}{b}}\right] }
 {1-{\exp}\left[{-\frac{\Delta}{b}}\right] }\,
 \frac{ b\,\sin\Delta x } {\sinh \pi b\,x}~.
 \label{S5}
\end{eqnarray}
 Together with the overall decrease of the cross section
 proportional to $\exp[-2\pi b\,l(\omega')]$ it also
 indicates fast oscillations
 with a frequency $\propto \Delta$. Such oscillations show up
 in the cross section as some secondary bumpy structures. These properties
are manifest in Fig.~\ref{Fig:4}~(bottom panel):
 the overall decrease of the cross section decreases with
 decreasing pulse duration, and the number of the secondary
 bumps in the region of $\omega'$, corresponding to the nearest
 integer values of $l$,  increases with pulse duration.

 \begin{figure}[h!]
\includegraphics[width=0.95\columnwidth]{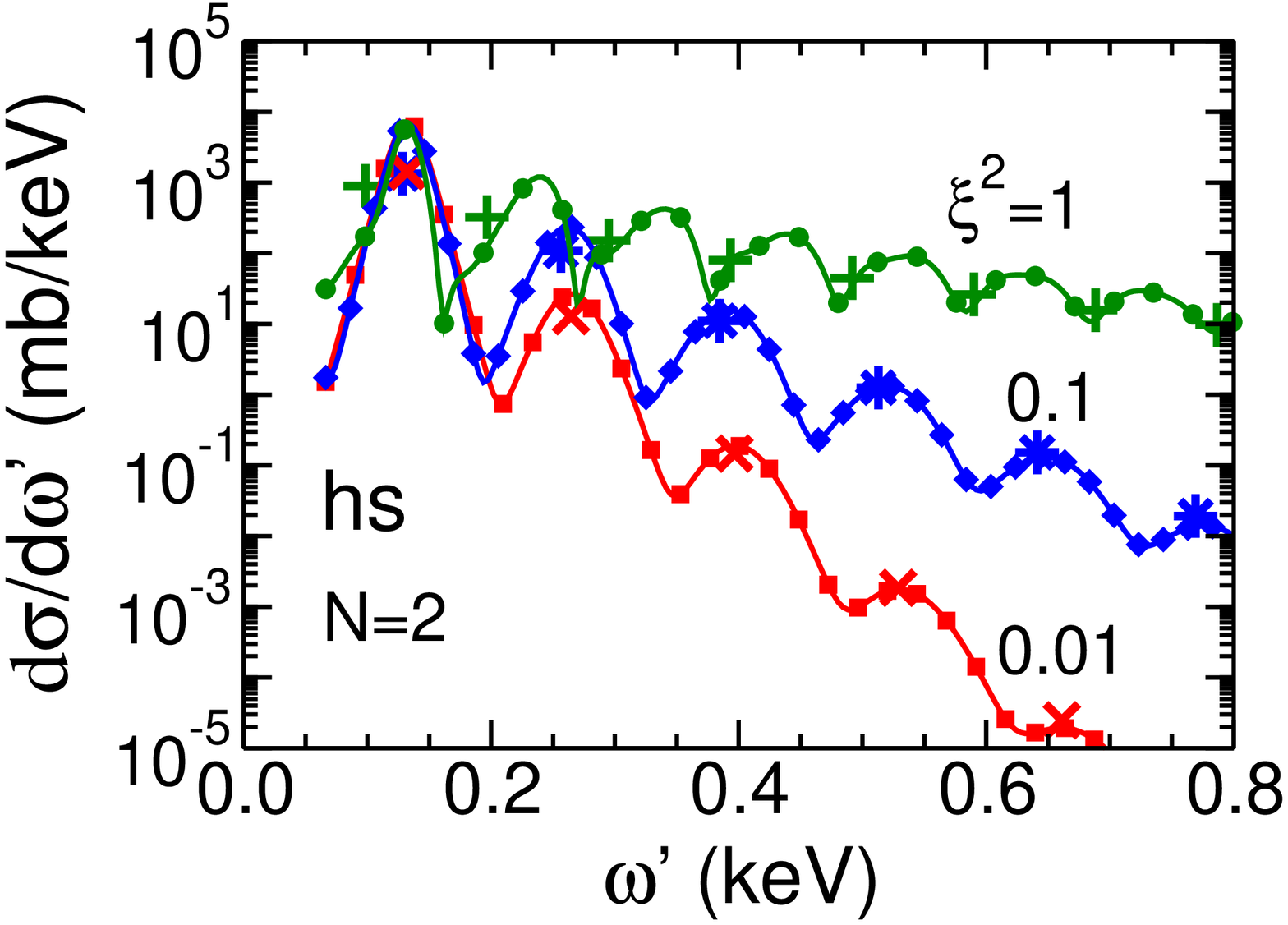}\qquad
\includegraphics[width=0.95\columnwidth]{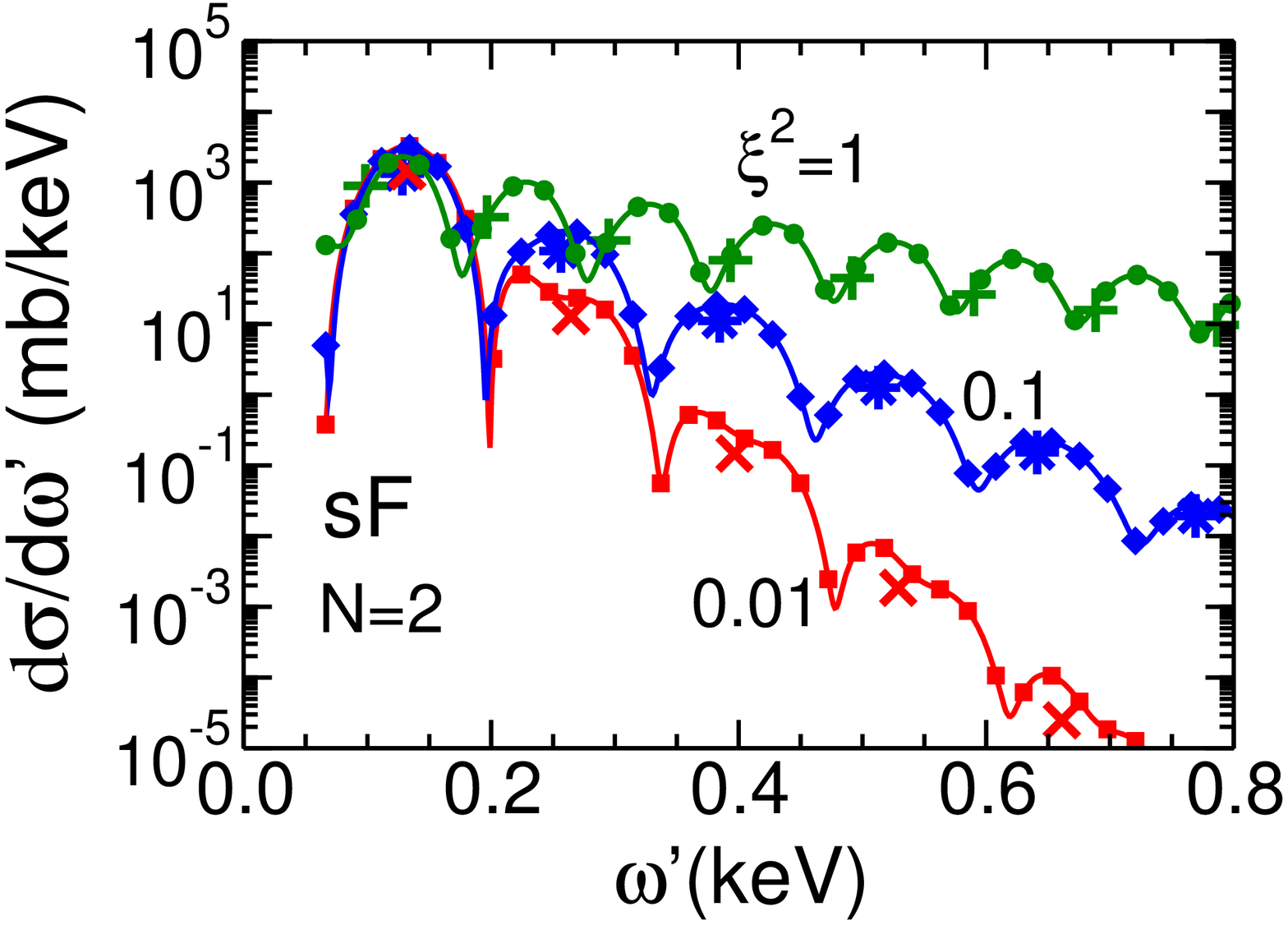}
 \caption{\small(Color~online) Differential cross section $d \sigma / d\omega' \,
 \vert_{\theta' = 170^o}$ of Compton scattering for
 $\xi^2=0.01,\,0.1$ and 1, shown
 by red (marked by boxes), blue (marked by diamonds),
 and green (marked by circles)curves,
 respectively, for $N=2$.
 The symbols "x", stars and pluses  depict the IPA results
 for the lowest harmonics for $\xi^2= 0.01,\,0.1,$ and 1, respectively.
 Top and bottom panels
 correspond to hyperbolic secant (hs) and symmetrized Fermi (sF)
 shapes of the envelopes.
%In the latter case, $b/\Delta=0.15$.
 \label{Fig:5}}
\end{figure}
 In Fig.~\ref{Fig:5} we present the
 differential cross sections
 for different field intensities $\xi^2=0.01,\,0.1$ and 1,
 depicted by red (marked by boxes), blue (marked by diamonds),
 and green (marked by circles) curves, respectively.
The duration of the pulse corresponds to $N=2$. The bump positions
 for FPA in Fig.~\ref{Fig:11} are shifted relative to the discrete
 positions of contributions from the individual harmonics in IPA,
 shown by corresponding symbols.
 These shifts are a consequence of the electron dressing in IPA
 which depends on $\xi^2$.

\begin{figure}[h!]
\includegraphics[width=0.95\columnwidth]{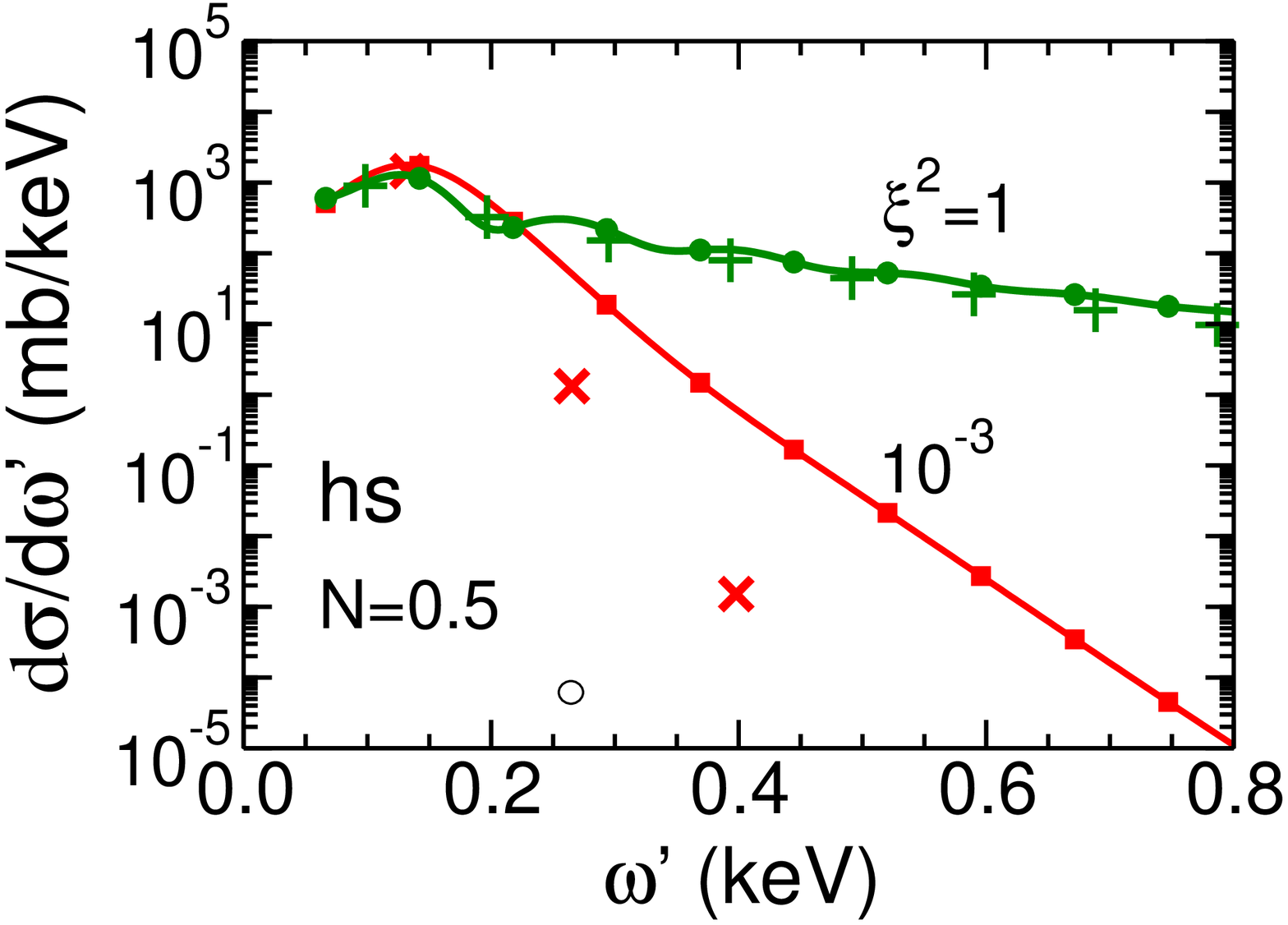}\qquad
\includegraphics[width=0.95\columnwidth]{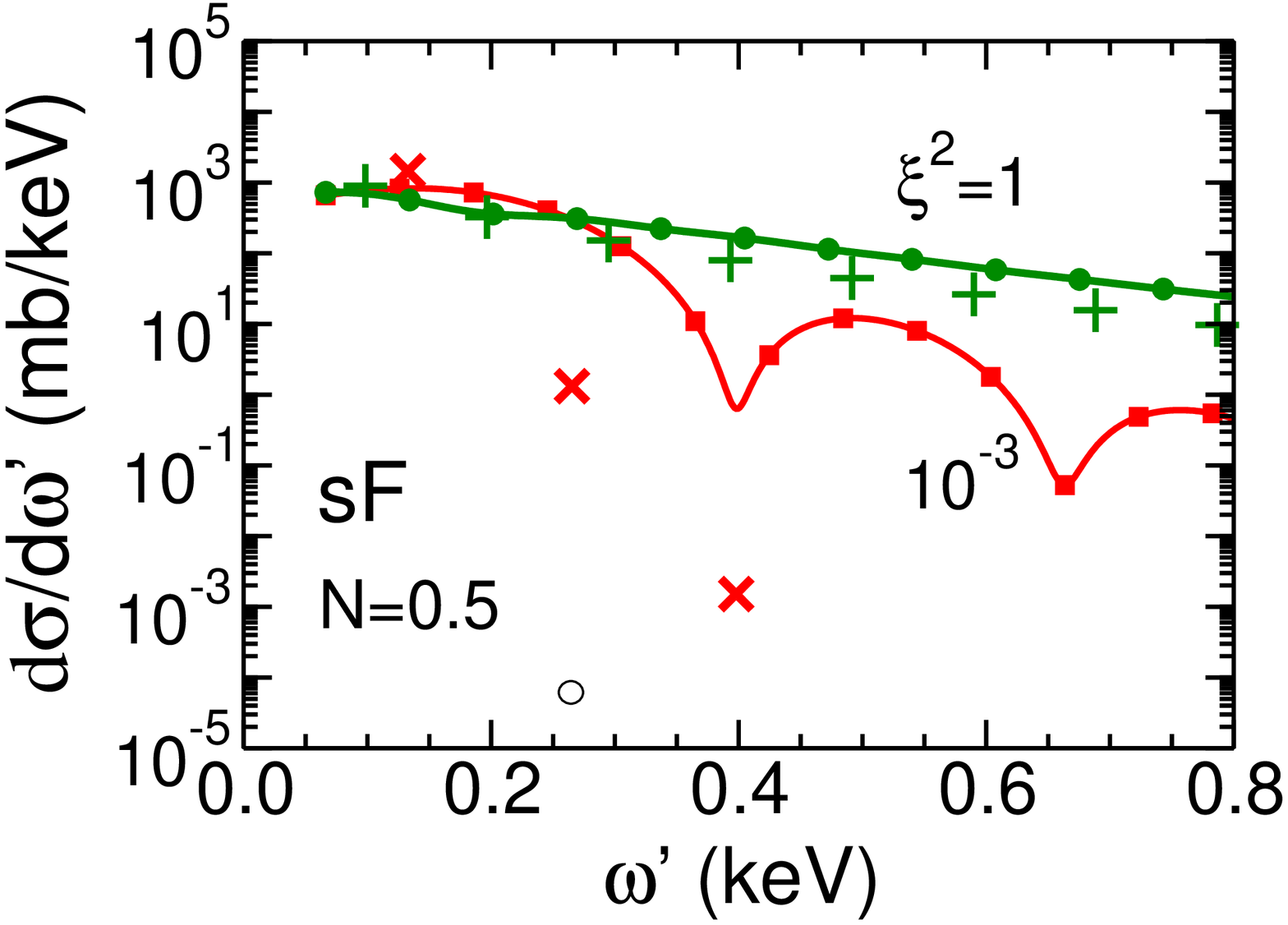}
 \caption{\small(Color~online) Differential cross section
 $d \sigma / d\omega' \,
 \vert_{\theta' = 170^o}$ of Compton scattering for
 $\xi^2=10^{-3}$ and 1 shown by red (with boxes)
 and  green (with circles) curves, respectively, for $N=0.5$.
 Crosses and pluses depict the discrete IPA results
 for lowest harmonics
 for $\xi^2=10^{-3}$ and 1, respectively.
 Top and bottom panels
 correspond to hyperbolic secant (hs) and symmetrized Fermi (sF)
 envelope shapes.
%In the latter case, $b/\Delta=0.15$.
 \label{Fig:6}}
\end{figure}

 For completeness, in Fig.~\ref{Fig:6} we exhibit
 the differential cross sections for a sub-cycle pulse with $N=0.5$
 for $\xi^2=0.001$ and 1, shown by red curves (marked by boxes)
 and green curves (marked by circles), respectively,
 for the hs (top) and sF (bottom) envelope shapes, respectively.
 Crosses and pluses depict the IPA results
 for $\xi^2=0.001$, and 1.
For the hs shape, the cross sections decrease almost monotonically,
with a large enhancement of the FPA result compared to IPA for
small field intensities ($\xi^2\ll 1$). In case of the flat-top
envelope the cross section exhibits some oscillations which
point to more complicated spectral properties of the flat-top
envelope shape.

 To summarize this part we can conclude that
 predictions for fully differential cross sections
 for IPA and FPA are quite different. In IPA, the cross section
 represents the discrete spectrum where the frequencies of the
 outgoing photons $\omega'$ are fixed according to Eq.~(\ref{S3_}).
 The fully differential cross sections are continuous
 functions of $\omega'$. Some similarities of IPA and FPA
 can be seen in the case of small field intensities $\xi^2\ll1$ and
the smooth one-parameter envelope shape with $N=2\dots10$. Here,
 the differential cross sections
 have a bump structure, where the position of bumps
 and bump heights are close to that predicted by IPA.
 The situation changes drastically for more complicated
 (and probably more realistic) flat-top envelope shapes. In this case
 one can see a lot of additional bumps which reflect the more
 complicated spectral properties of the flat-top  shape;
% In this case
it is difficult to find a relation not only between IPA and
 FPA, but also within FPA for different pulse durations.
 Experimentally studying multi-photon effects using rapidly
 oscillating fully differential cross sections seems
 to be rather complicated. An
 analysis of integral observables helps to overcome this problem.
 In particular, the partly integrated cross sections have a
 distinct advantage: they are smooth functions of $\omega'$ and
 allow to study directly the multi-photon dynamics.

\subsection{Partly integrated cross sections}

%%%%%%%%%%%%%%%%%%%%%%%%%%%%%%%%%%%%%%%%%%%%%%%%%%%%%%%%%%%%%%%%%
 Non-linear effects become most transparent  in the
 partially energy-integrated cross section defined in Eq.~(\ref{S6}).
 In this case, the sub-threshold multi-photon events are filtered
 when the lower limit of integration $\omega'$ exceeds the
 threshold value $\omega_1'=\omega'(n=1,\xi^2)$ (with $\xi^2\ll1$
 for the pure Klein-Nishina process). Thus, events with
 $\omega'(l)\gg\omega'_1$ and $l\gg1$ correspond essentially to
 multi-photon process, where the energy $l\omega\gg\omega$ is
 absorbed from the pulse.
 Experimentally, this can be realized by an absorptive medium
 which is transparent for frequencies above a certain threshold
 $\omega'$. Otherwise, such a partially integrated spectrum can be
 synthesized from a completely measured spectrum. Admittedly, the
 considered range of energies with a spectral distribution
 uncovering many decades is experimentally challenging.

 \begin{figure}[h!]
\includegraphics[width=0.95\columnwidth]{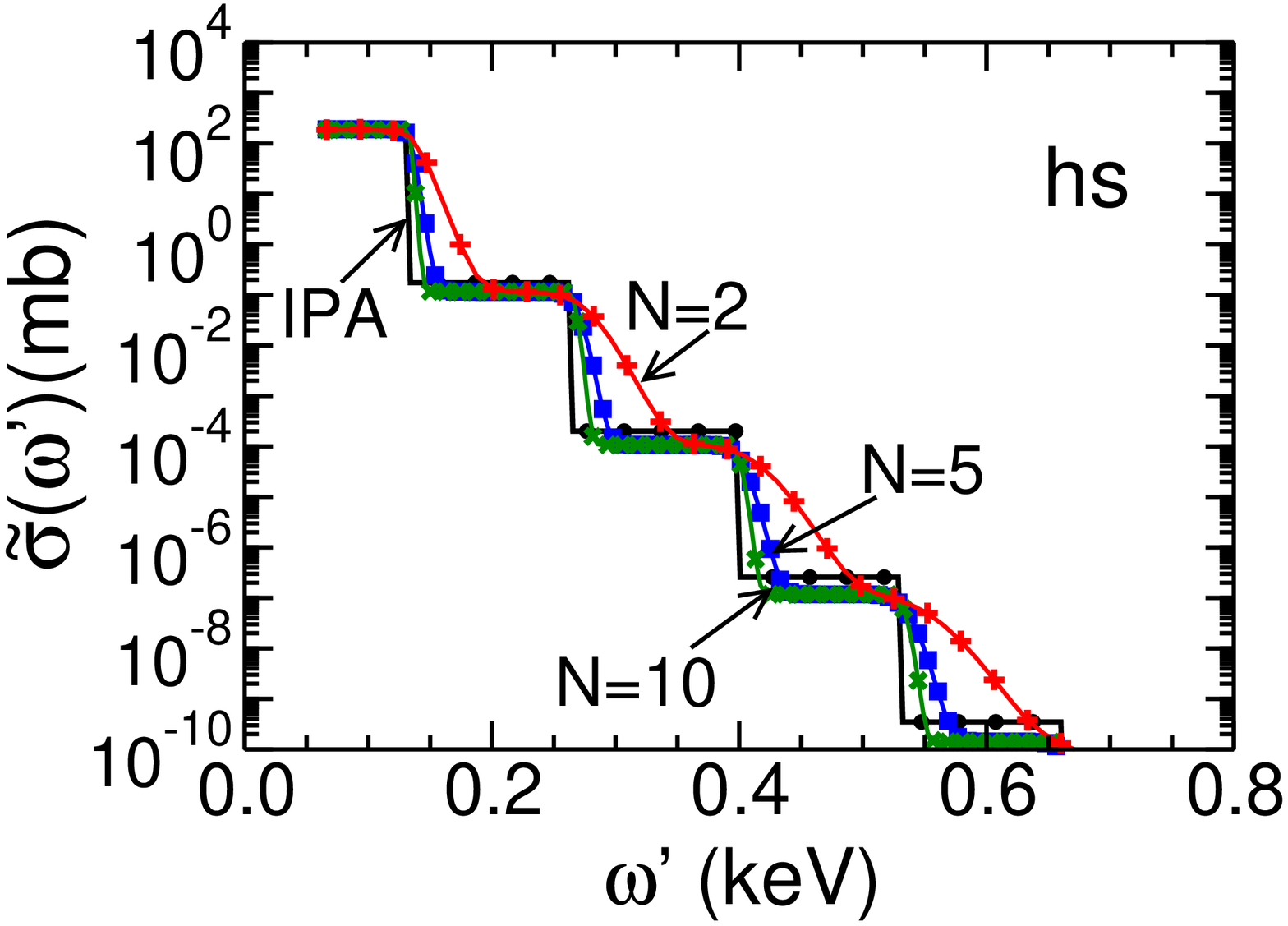}\qquad
\includegraphics[width=0.95\columnwidth]{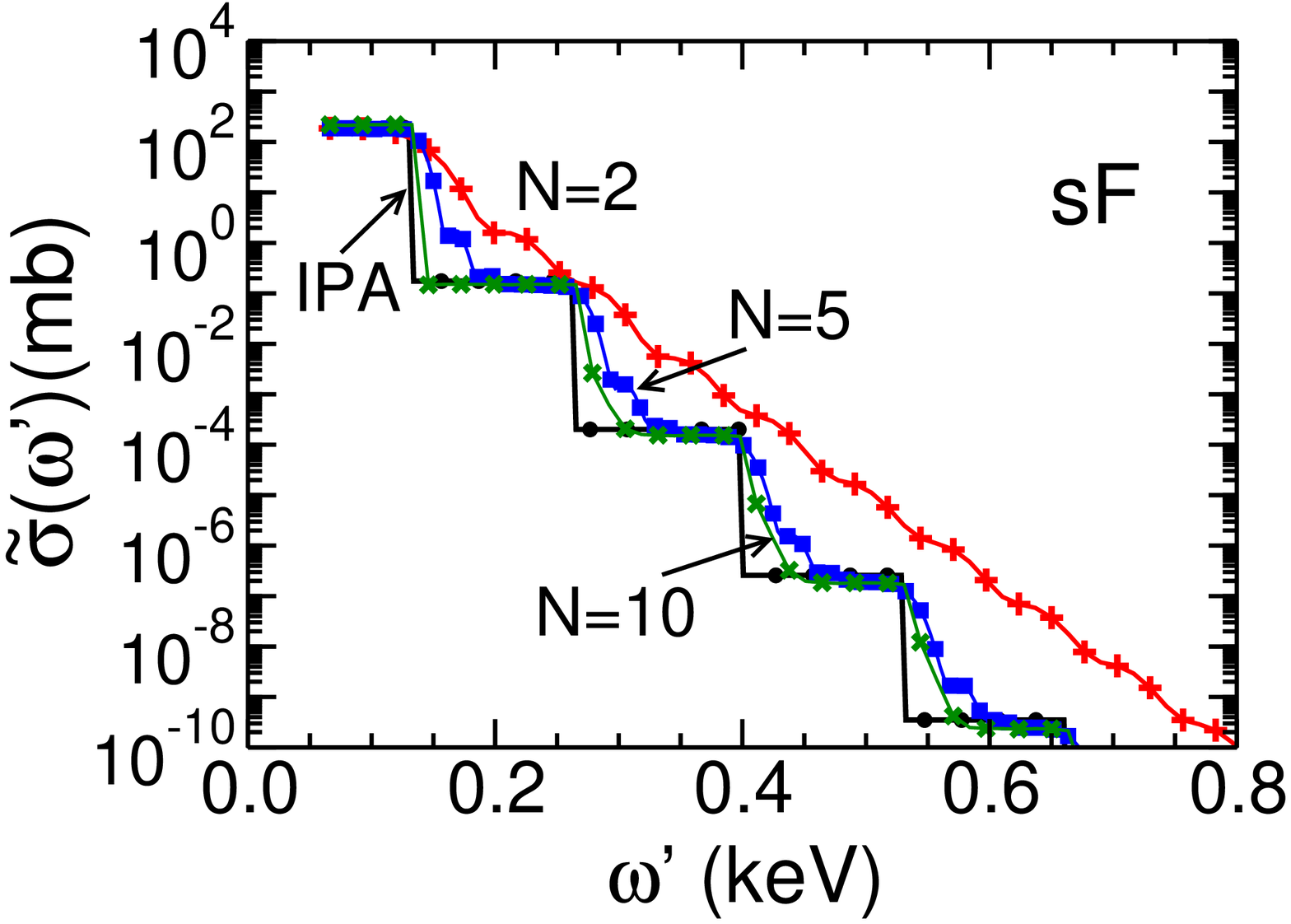}
 \caption{\small{(Color online) The integrated cross section (\ref{S6})
 for $\xi^2=10^{-3}$.
 The thin solid curve marked by dots depicts the IPA result.
 The red (with pluses), blue (with boxes) and green (with crosses) curves
 correspond to $N =2$, 5 and 10, respectively. Top and bottom
 panels are for hyperbolic secant (hs) and symmetrized Fermi (sF) envelopes.
% In latter the case, $b/\Delta=0.15$.
\label{Fig:7} }}
\end{figure}
 The partially integrated cross sections of Eq.~(\ref{S6})
 are presented in Fig.~\ref{Fig:7}.
 The thin solid black curve (marked by dots) depicts IPA results given by
\begin{eqnarray}
\tilde\sigma^{IPA} (\omega')
 =\int\limits_{l'(\omega')}^{\infty}dl\sum\limits_{n=1}^{\infty}
 \frac{d\sigma^{IPA}_n}{d\omega'_n}
 \frac{d\omega'_n}{d n}\theta(n-l)~,
 \label{S6_IPA}
\end{eqnarray}
% with
% \begin{eqnarray}
% \frac{d\omega' (n)}{dn}=
% \frac{{\omega'}^2(E+|\vec p|\cos\theta' +\omega^2\xi^2(1-\cos\theta'/2(k\cdot
% p)))}
% {n^2\omega(E+|\vec p|)}~,
%\label{S66_IPA}
% \end{eqnarray}
 where $\omega'(n)$ is defined by Eq.~(\ref{S3_}).
 That is, the partially integrated cross section becomes a step-like
 function, where  each new step corresponds to the contribution of
 a new (higher) harmonic $n$,
 which can be interpreted  as $n$-laser photon process. Results for
 the finite pulse
 exhibited by red (marked by pluses), blue (marked by boxes)
 and green (marked by crosses) curves
 correspond to $N=2, \,5$ and
 10, respectively. In the above-threshold region with
 $\omega'\leq \omega'_1$, the cross sections do not depend
 on the widths and shapes of the envelopes, and the results of IPA and FPA
 coincide. The situation changes drastically
 in the deep sub-threshold region, where $\omega'>\omega_1'$ $(l\gg1),\, n\gg1$.
 For the short pulses with $N\simeq 2$,
 the FPA results exceed that of IPA considerably,
 and the excess may reach
 several orders of magnitude, especially for the flat-top envelope
 shown by the red curve in Fig.~\ref{Fig:7}~(bottom panel).
 However, when the number of oscillation in a pulse
 increases ($N\gtrsim 10$)
 there is  a qualitative convergence
 of FPA and IPA results, independently of the pulse shape.
 Thus, at $N=10$ and $\omega'=0.6$~keV the difference between predictions
 for hs and sF shapes is a factor of two, as compared with
 the difference of the few orders of
 magnitude at $N=2$ for the same value of $\omega'$.

\begin{figure}[h!]
 \includegraphics[width=0.95\columnwidth]{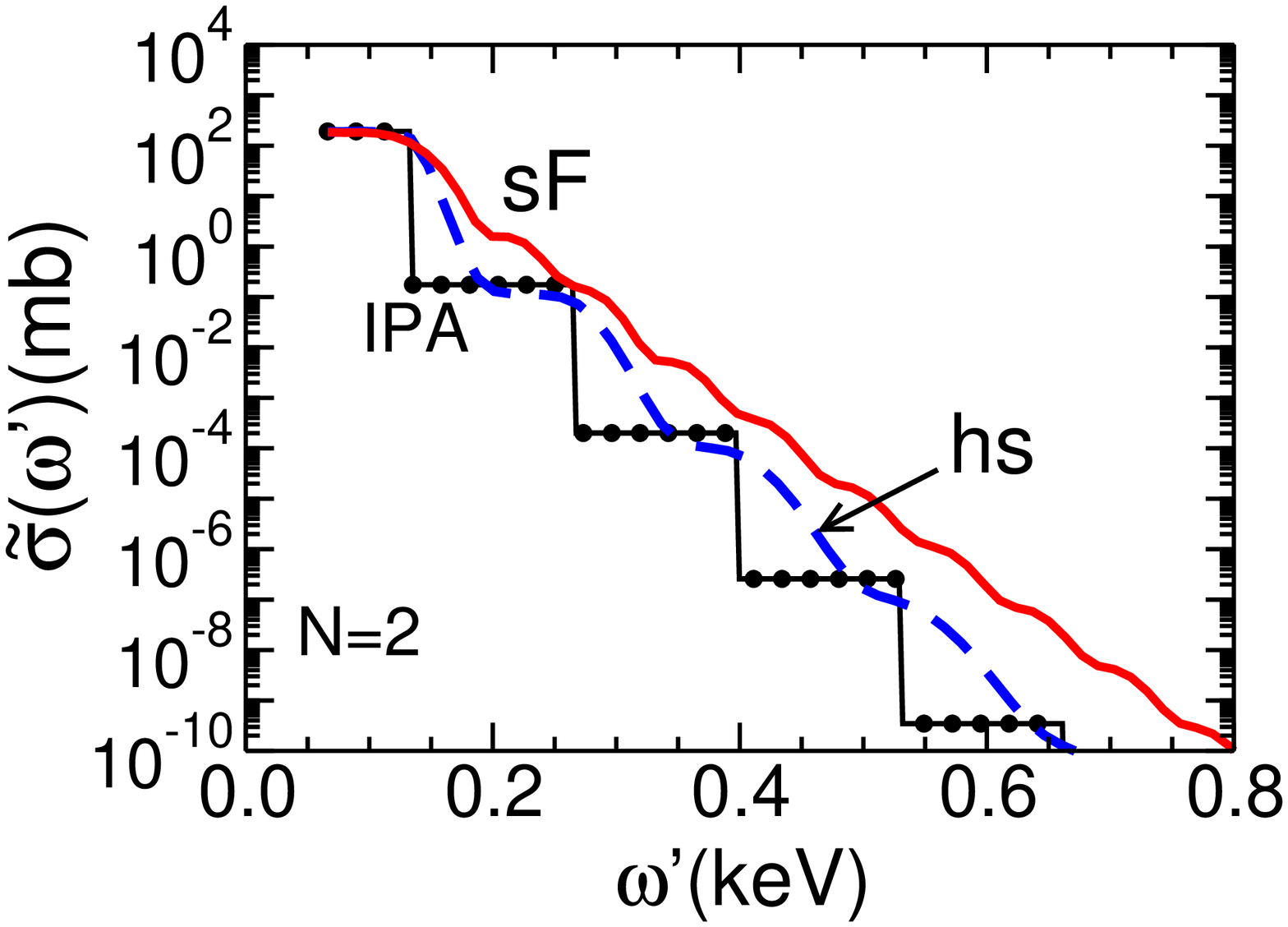}
 \includegraphics[width=0.95\columnwidth]{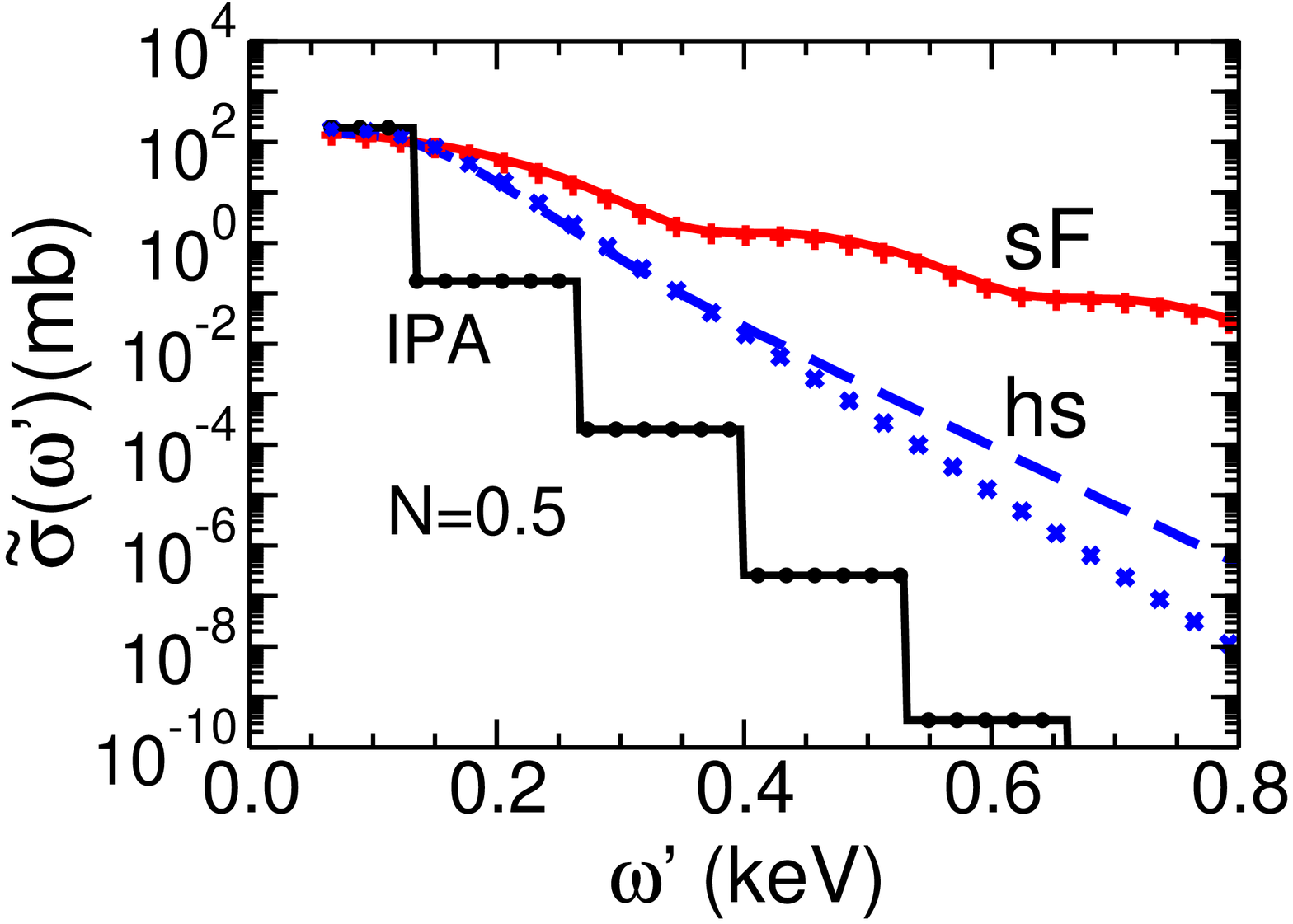}
 \caption{\small{(Color online)
 The partially integrated cross section (\ref{S6}) for $\xi^2=10^{-3}$.
 Top  panel: $N=2$, for the hyperbolic secant (hs, dashed blue curve)
 and symmetrized Fermi (sF, solid red curve) shapes.
 Bottom panel: The same as in top panel, but for a sub-cycle pulse
 with $N=0.5$.
 The crosses and pluses correspond to the asymptotic solutions
 for hs and sF shapes, respectively, described in
 the text.
 \label{Fig:8} }}
\end{figure}

 To highlight the difference of the hs and
 sF (flat-top) shapes for short pulse we exhibit in Fig.~\ref{Fig:8}
 (top panel) results for $N=2$.
 At $l \gtrsim5$, corresponding to $\omega' \gtrsim 0.7$~keV,
 the difference between them is more than two orders of magnitude.

 Consider now the case of sub-cycle pulses
 with $N<1$. Our result for $N=0.5$ %and small field intensity ($\xi^2=10^{-3}$)
 is exhibited in Fig.~\ref{Fig:8}~(bottom panel).
 One can see a large enhancement
 of the cross section with respect to the IPA case
 for the sub-cycle pulse in the sub-threshold region.
 The enhancement for the sF shape is much greater
 pointing to a sensible dependence on the actual pulse shape.
 For a qualitative estimate of such a behavior we can drop
 the $\phi_{e'}$ dependence by taking $\phi_{e'}=0$.
 This choice is quite reasonable for the flat-top
 sF envelope shape and may serve as an upper limit
 for the cross sections in the case of the smooth
 hs envelope shape (cf.\ Sect.~III.D).
 Under the considered
 conditions the basic function $Y_l$ in Eq.~(\ref{B6})
 can be approximated as
 \begin{eqnarray}
  Y_l&\simeq& \frac{1}{2\pi}\int dq\,F(q)\int d\phi
  {\rm e}^{i(l-q)\phi -i{\cal P}(\phi)}\nonumber\\
  &\simeq&\frac{1}{2\pi}\int dq\,F(q)\int d\phi
  {\rm e}^{i(l-q-l\beta\xi)\phi -i\delta}\nonumber\\
  &=&{\rm e}^{-i\delta} F(\tilde l)~,
 \label{S_FF}
 \end{eqnarray}
 where $F(l)$ is the Fourier transform of the envelope
 function,
 $\tilde l=l(1-\beta\xi)$ with
 $\beta=2\sqrt{\frac{u}{u_l}(1-\frac{u}{u_l})}<1$
 and $\delta=z\int_{-\infty}^0 d\phi \cos\phi\,f(\phi)$.
 As a result, the cross section is almost completely defined
 by the square of the Fourier transforms
 (cf.\  Eqs.~(\ref{S4}) and (\ref{S5})), i.e.
 $\tilde\sigma(\omega')\simeq
 g(l(\omega'))\,F^2(\tilde l(\omega')-1 )$,
 where $g(\omega')$ is a smooth function of $l=l(\omega')$
 (cf. Eq.~(\ref{ASY4})).
 The Fourier transform for the sF shape decreases slower with
 increasing $l$.
 Such a dependence is evident in Fig.~\ref{Fig:7} (bottom panel).
 For an illustration,
 the crosses depict the result of a calculation where
 the basic functions $Y_l$ and $X_l$
 in the partial probability $\omega' (l)$ in Eq.~(\ref{S2})
 are replaced by their asymptotic values $F^{(1)}(\tilde l-1)$
 and $F^{(2)}(\tilde l-1)$. A more detail discussion of
 the asymptotic result is presented below (cf.\ Eq.~(\ref{ASY4})).
% One recognizes a qualitative agreement between full
% and asymptotic results, in particular for the sF envelope.
% This result can be used easily for a qualitative analysis.

\begin{figure}[h!]
 \includegraphics[width=0.95\columnwidth]{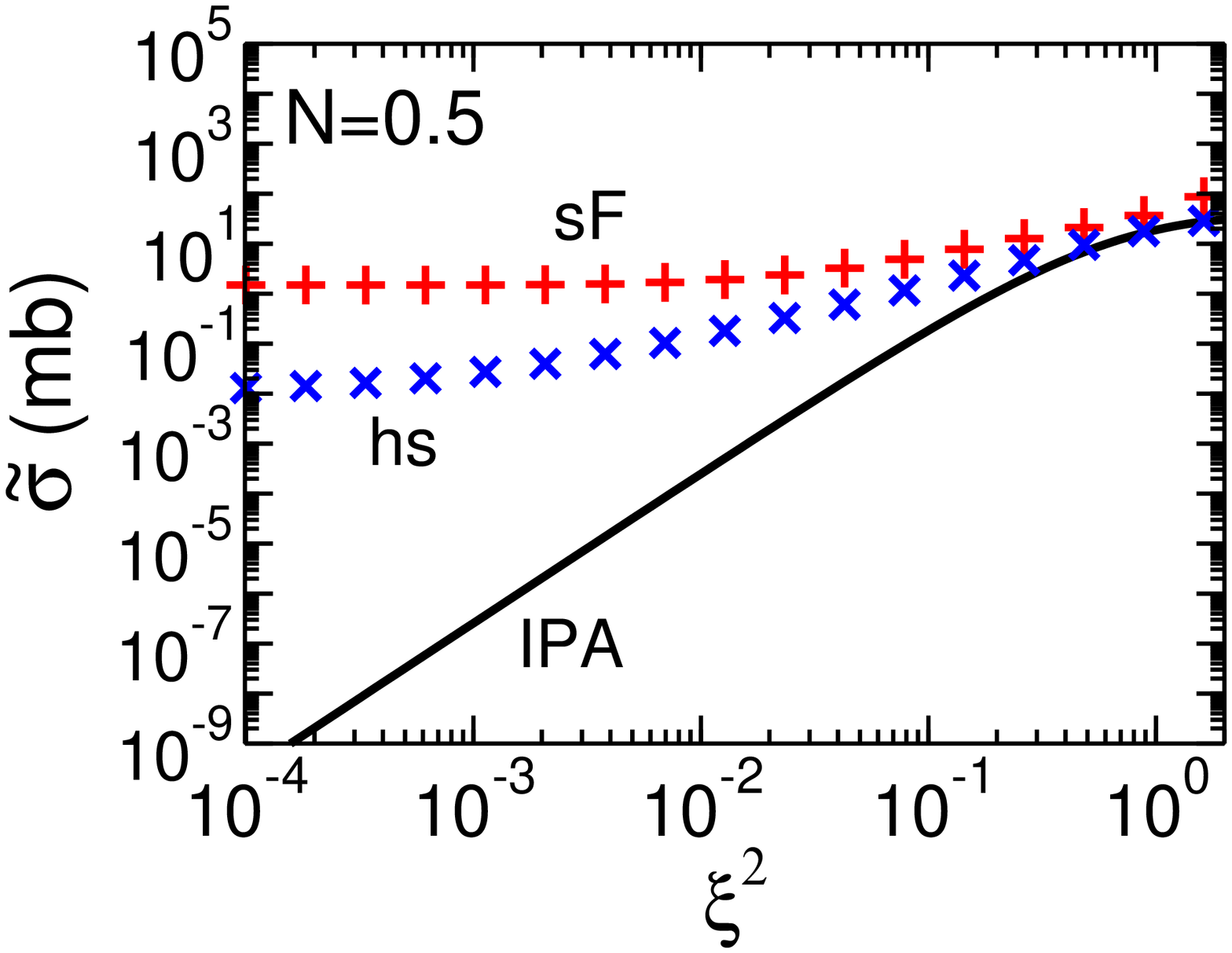}\\
 \vspace*{1mm}
 \includegraphics[width=0.95\columnwidth]{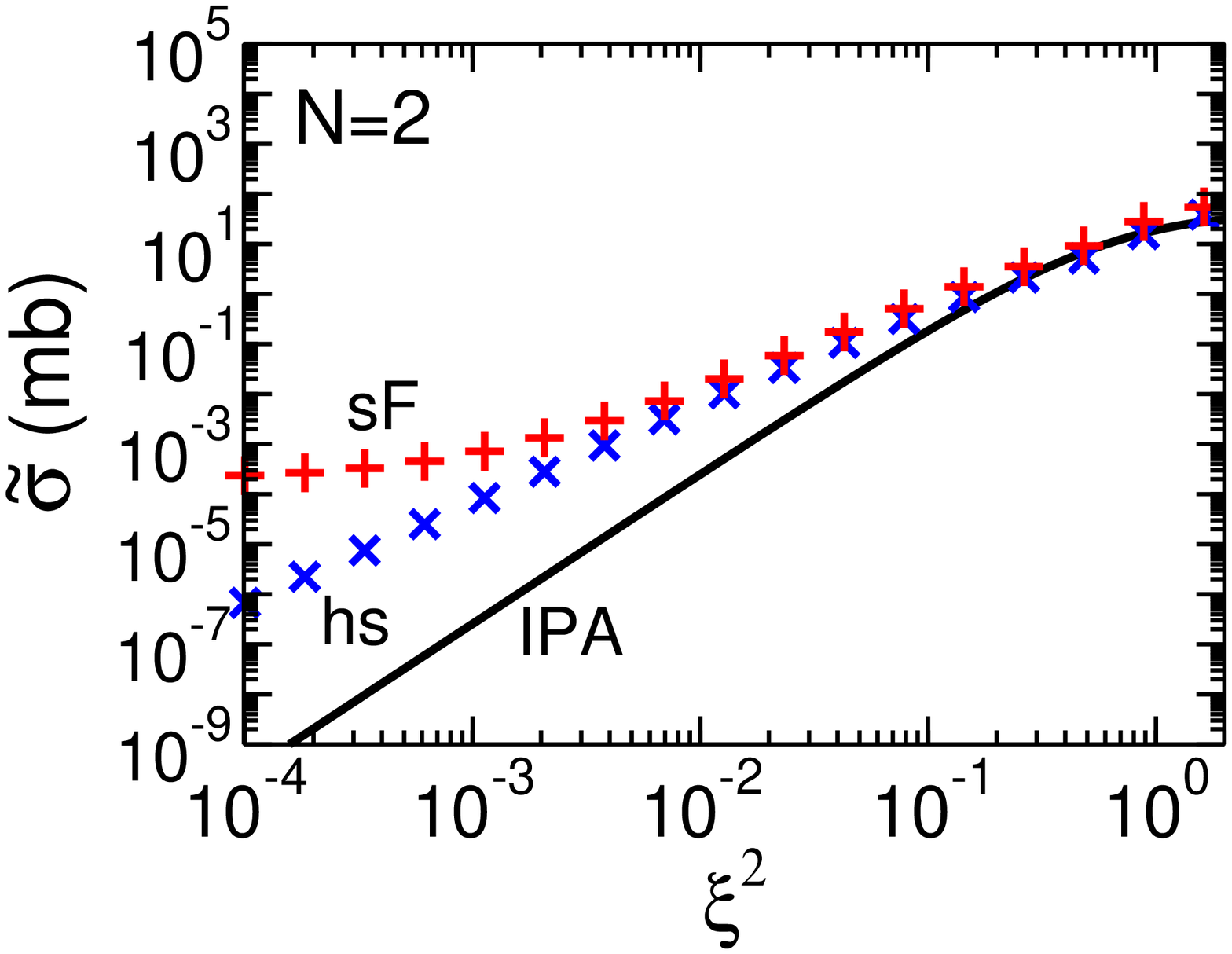}
 \caption{\small{(Color~online)~The partially integrated cross
 section as a function of $\xi^2$ at $\omega'/\omega_1=3$
 for short pulses with $N=0.5$ (top) and 2 (bottom).
 The solid curve and symbols correspond to IPA and FPA
 (hs and sF envelope functions), respectively.
% In case of sF shape $b/\Delta=0.15$.
\label{Fig:9}}}
 \end{figure}

  The dependence of the partially integrated cross
  section as a function of $\xi^2$ at fixed ratio $r\equiv\omega'/\omega'_1=3$
  for short pulses with $N=0.5$ and 2 is exhibited in Fig.~\ref{Fig:9}
  in top and bottom panels, respectively.
  Note that the minimum value of $l'(\omega')$
  is related to $r$ as
  \begin{eqnarray}
 l'(\omega')&=&r\frac{E+|\vec p|\cos\theta'}
 {E+|\vec p|\cos\theta'+\omega(1-r)(1-\cos\theta')}~,
 \label{S_Lim_l}
 \end{eqnarray}
 meaning $l' < r$. Similarly, for $n_{\rm min}$ one has
 $n_{\rm min}=x$, for $I(x)$=x and  $n_{\rm min}=x+1$
 for $I(x)<x$ with
\begin{eqnarray}
 x =\frac{E+|\vec p|\cos\theta'
 +\frac{\omega m^2\xi^2}{2(k\cdot p)}(1-\cos\theta')}
 {E+|\vec p|\cos\theta'+\omega(1-r
 + \frac{m^2\xi^2}{2(k\cdot p)})(1-\cos\theta')} .
 \label{S_Lim}
 \end{eqnarray}
 The solid curves and symbols correspond to IPA and FPA,
  respectively, with different pulse shapes.
 One can see that the main difference of
 IPA and FPA, as well as the pulse shape dependence,
 appears at small field intensities
 $\xi^2\ll1$, where the dependence of the cross section on
 the pulse shape and duration is essential.

 To explain this result we use
 the asymptotic solution for $\tilde \sigma$ which is obtained
 by keeping leading terms in $\xi^2$ in Eqs.~(\ref{S2}) and (\ref{S2INF})
 and taking into account
 that the dominant contribution to the integrals of
 Eqs.~(\ref{S6}) and (\ref{S6_IPA}) stems
 from $l\sim l'$ and $n\sim I(l')+1$, respectively.
 Consider first the partially integrated cross section in IPA.
 Using the asymptotical expression for the Bessel
 functions
 \begin{eqnarray}
 J_k(z)\simeq \left(\frac{z}{2}\right)^k\frac{1}{k!} \quad
{\rm for}\,\,\,z\ll1~,
 \label{ASY1}
 \end{eqnarray}
 and keeping the leading terms in Eq.~(\ref{S2INF}) with
 $J_{n-1}^2(z)$ and $n=I(l')+1$, one obtains
 \begin{eqnarray}
 \tilde\sigma^{IPA}\simeq
 \frac{2\pi\alpha^2}{(E+|\vec p|\cos\theta')|\vec p|}\xi^{2k}
 \Phi(k)~,% + {\cal O}{\xi^{2(k+1)}}~,
 \label{ASY2}
 \end{eqnarray}
 where $k=I(l')\simeq I(r)$ and
  \begin{eqnarray}
 \Phi(k)&=&\frac{(k+1)^{2(k+1)}}{(k+1)!^2}
 \left(t_k(1-t_k)\right)^{2k}\nonumber\\
 &\times&\left(1+\frac{u}{2(1+u)}-2t_k(1-t_k)\right)
 \label{ASY22}
 \end{eqnarray}
 with $t_k=u/u_k$, where
 $u=\omega'(1-\cos\theta')/(E+|\vec p|-\omega'(1- \cos\theta') )$
 and $u_k=2k\omega(E+|\vec p|)/m^2$. Within the considered kinematics,
 $t_k$ does not depend on $k$ and can be approximated by
 $t_k\simeq m^2(1-\cos\theta')/(2(E+|p|\cos\theta')(E+|p|))\simeq
 0.35$.
\begin{figure}[h!]
 \includegraphics[width=0.8\columnwidth]{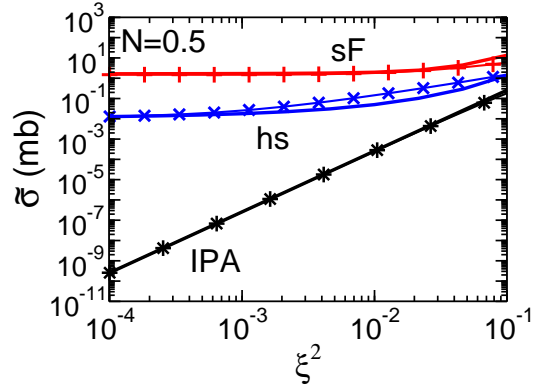}
 \caption{\small{(Color~online)~
 The partially integrated cross
 sections as a function of $\xi^2\ll1$ for sub-cycle pulse
 with $N=0.5$.
 The stars are for the full IPA result. The black curves correspond to the
 asymptotic solution of Eq.~(\ref{ASY2}).
 The red and blue thin curves marked by
 pluses and crosses  are full calculations
 for sF and hs shapes, respectively,
 while the thick red and blue curves
 are the corresponding asymptotic results
 of Eq.~(\ref{ASY4}).
% In case of sF shape $b/\Delta=0.15$.
 \label{Fig:10}} }
 \end{figure}

 The result for asymptotic solution of (\ref{ASY2})
 is shown by
 the solid black curve in Fig.~\ref{Fig:10} together with a full
 calculation depicted by stars. One can see an excellent agreement
 of these two results.

 For FPA in case of sub-cycle pulse with $N=0.5$,
 we use the asymptotic
 representation for the basic functions $Y_l$ in the form of
 Eq.~(\ref{S_FF}) which allows to express
 the partially integrated cross section as
  \begin{eqnarray}
  &&\tilde\sigma\simeq
  \frac{2\pi\alpha^2}{N_0(E+|\vec p|\cos\theta')|\vec p|}\nonumber\\
  &\times& \left(1+\frac{u}{2(1+u)}-2t_{l'}(1-t_{l'})\right)
  \int\limits_{l'}^{l'+1} dl\,F^2(\tilde l-1)~,
  \nonumber\\
  \label{ASY4}
  \end{eqnarray}
 where $F(x)$ is the Fourier transform of the envelope
 function (cf. Eqs.~(\ref{S4}) and (\ref{S5})).
 Results for the sub-cycle pulse with $N=0.5$ are
 presented in Fig.~\ref{Fig:10}, where
 the red and blue thin curves marked by
 pluses and crosses  are for full calculations
 for the sF and hs shapes, respectively.
 The red and blue thick curves are the asymptotic
 results of Eq.~(\ref{ASY4})
 for sF and hs shapes, respectively.

  \begin{figure}[h!]
 \includegraphics[width=0.75\columnwidth]{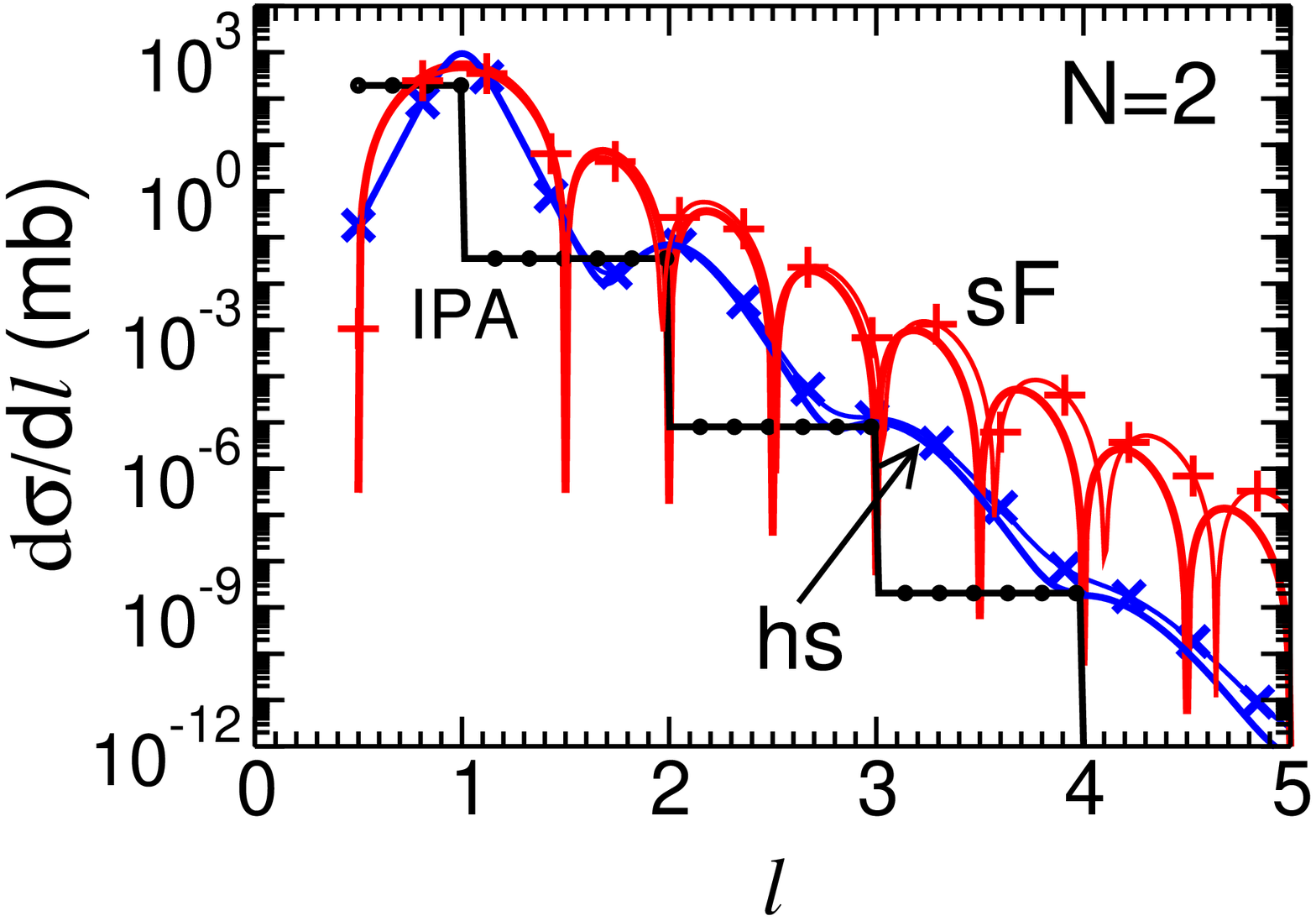}
 \includegraphics[width=0.8\columnwidth] {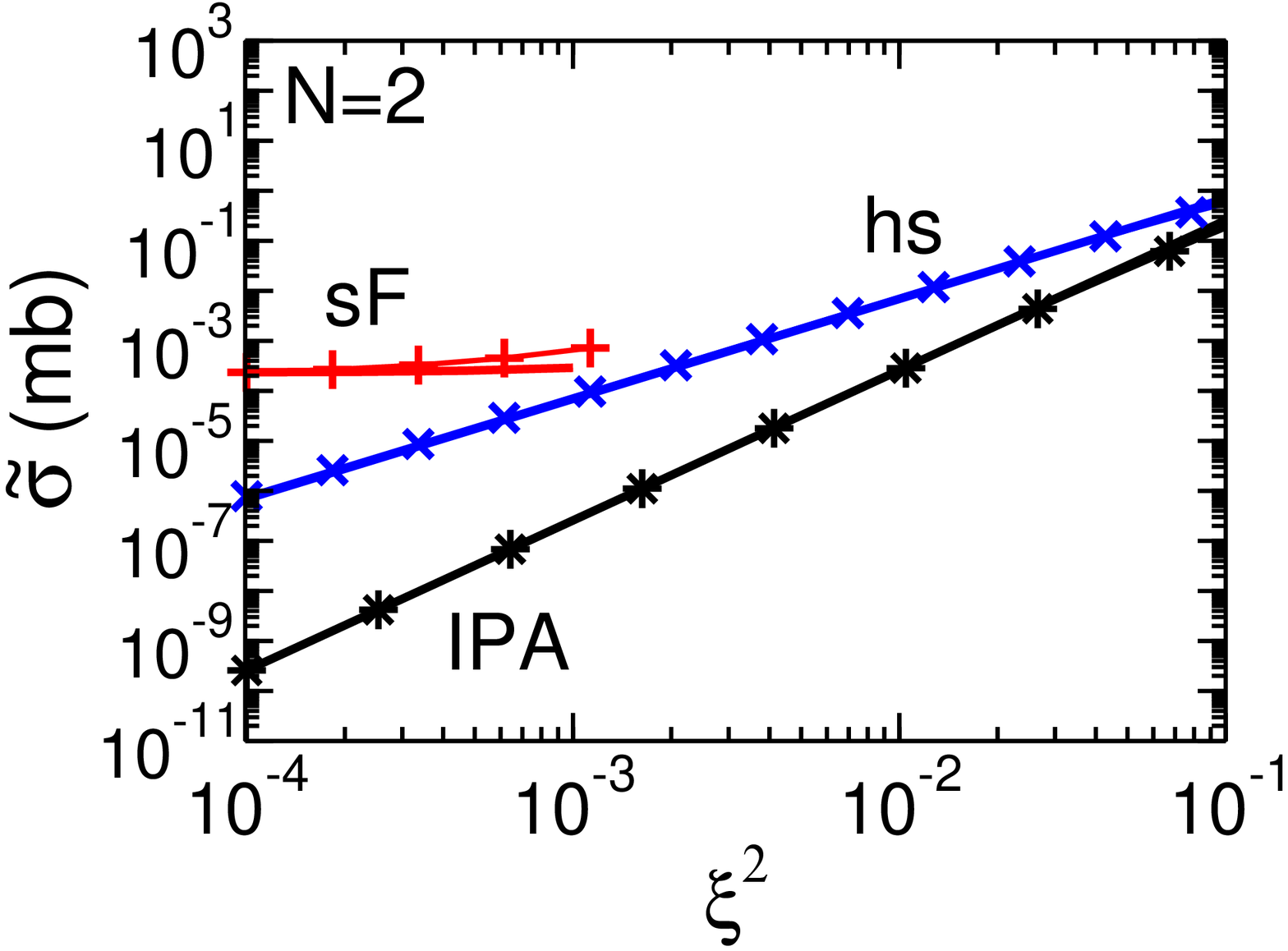}
 \caption{\small{(Color~online)
 Results for a short pulse with $N=2$.
 Top panel: The integrand ${d\sigma}/{dl}$  of the partially
 integrated cross sections for $\xi^2=2\times 10^{-4}$.
 The thin black (with circles) blue
 (with crosses) and red (with pluses) curves are for the full
 calculations. The asymptotic predictions
 (described in the text) are exhibited
 by the thick black, blue and red curves, respectively.
 Bottom panel:
 The black, red and blue thin curves marked by
 stars, pluses and crosses are full calculation,
 for IPA and FPA for sF and hs shapes, respectively;
 the thick black, red and blue curves
 are the corresponding asymptotic results
 of Eqs.~(\ref{ASY2}), (\ref{ASY4})
 and (\ref{ASY7}), respectively.
% In case of sF shape $b/\Delta=0.15$.
 \label{Fig:11}} }
 \end{figure}

  We would like to note that, at $\xi^2\ll1$, our asymptotic solution
  for sub-cycle pulse weakly depends on $\xi$ only through
  the weak $l(1-\beta\xi)$ dependence in the Fourier transform.
  The leading $\xi^2$
  dependence of the partial harmonics $w_l$ in (\ref{S2}) is
  compensated by the $\xi^2$ dependence of the flux factor in the
  denominator of Eq.~(\ref{S1}). Nevertheless, such a weak $\xi$ dependence
  is in qualitative agreement  between full and
  asymptotic solutions both for sF and hs envelope shapes.
  Thus, we can conclude that the partially integrated
  cross section for the sub-cycle pulse at $\xi^2\ll1$ is almost completely
  determined by the square of the Fourier transform of the envelope function
  which is a measure of high momentum frequencies
  generated by the pulse shape.

  In the case of a short pulse with $N=2$ and $\xi^2=2\times 10^{-4}$,
  the integrand of $\tilde\sigma$ for the hs shape
  for the full calculation is shown by the blue thin
  curve marked by crosses in Fig.~\ref{Fig:11}, top panel.
  The integrand has a bump-like structure with the number of bumps
  equal to the number of partial harmonics in IPA,
  similarly to the differential cross section shown in
  Fig.~\ref{Fig:4}, top panel.
  For the asymptotic solution, we use in this case the
  asymptotic expression for the basic functions of Eq.~(\ref{B6}).
  Note that such an expression is valid only for the smooth
  one-parameter envelope shapes, where the function  ${\cal P}(\phi)$
  defined in Eq.~(\ref{III24})
  takes a simple form ${\cal P}(\phi)=z\sin(\phi-\phi_0)f(\phi)+{\cal
  O}(\xi^2)$~\cite{TKTH-2013}.
  One can see that, if the argument obeys $l' > I(l')$,
  then the main contribution to the cross section comes from
  the two harmonics with
 \begin{eqnarray}
 Y_{k,\varepsilon_1}(z)\,\,\,\,{\rm and}\,\,\,\,
 Y_{k+1,\varepsilon_2}(z)~,
 \label{ASY6}
 \end{eqnarray}
 where $k=I(l')$, $\varepsilon_1=l'-I(l')\equiv \varepsilon > 0$, and
 $\varepsilon_2=\varepsilon-1 < 0$.
 Then, keeping the leading terms in $\xi^2$ in (\ref{S2})
 one can get the approximate expression for the partly
 integrated cross section in the form
 \begin{eqnarray}
 && \tilde\sigma\simeq
 \frac{2\pi\alpha^2}{N_0(E+|\vec p|\cos\theta')|\vec p|}\xi^{2(k-1)}
 \nonumber\\
 &\times&\left(
 \Phi(k-1)\int\limits_{\varepsilon}^1\,d\epsilon \,(F^{(k)}(\epsilon))^2
 \right.
 \nonumber\\
 &&\left.\qquad\qquad +\xi^2\,\Phi(k)\int\limits_{\varepsilon-1}^1\,d\epsilon
  (F^{(k+1)}(\epsilon))^2
  \right) ,
 \label{ASY7}
 \end{eqnarray}
 where $F^{(m)}$ is the Fourier transform of $m$-th power
 of the envelope function $f(\phi)$. The integrand of the
 corresponding cross section is obtained from Eq.~(\ref{ASY7})
 by removing the integration  and putting $\varepsilon_1$ and
 $\varepsilon_2$ as the arguments of the squares of the first
 and second Fourier transforms, respectively.
 The approximate integrand is shown in Fig.~\ref{Fig:11},
  top panel, by the thick blue curve. One can see a reasonable
  agreement of the approximate and full calculations.
  The asymptotic expression of the integrand for IPA is equal
  $d\sigma_n/dl=\sigma_n\,\theta(n-l)$ with $\sigma_n$
  determined by Eq.~(\ref{ASY2}).

  The full and approximate results for $\tilde\sigma$ are shown in
  Fig.~(\ref{Fig:11}), bottom panel, by crosses and the
  thick blue curve, respectively. One can see a
  fairly good agreement of approximate and full results
  up to $\xi^2=0.1$.

  In the case of the flat-top envelope, the integrand of $\tilde\sigma$
  has a more complicated structure with a large number of bumps
  as shown in Fig.~\ref{Fig:11}, top panel,
  by the thin red curve marked by pluses.
  The asymptotic solution for the basic functions of Eq.~(\ref{B6})
  does not apply here. However, as a first approximation
  one can use the asymptotic solution of Eq.~(\ref{S_FF}). Then, the
  cross section $\tilde\sigma$ is determined by Eq.~(\ref{ASY4}).
  The asymptotic integrand is defined in this case by Eq.~(\ref{ASY4})
  by skipping the integration and putting $l=\hat l'$.
  The corresponding integrand is presented in  Fig.~\ref{Fig:11},
  top panel by thick red line. On can see a satisfactory agreement
  of full and approximate solutions and, as a consequence,
  a reasonable agreement of full and approximate calculations
  of partially integrated cross sections, however, in a very limited
  range of $\xi^2\ll1$ as shown in  Fig.~\ref{Fig:11},
  bottom panel, by pluses and the thick red line, respectively.

  To summarize this part we note that, in case of
  short pulses  and small field intensities, the partly
  integrated cross section is determined by
  the interplay of pulse shape and multi-photon dynamics.
  For both considered shapes, the cross sections are
  described by the simple asymptotic expressions
  which can be used in practical research.

%%%%%%%%%%%%%%%%%%%%%%%%%%%%%%%%%%%%%%%%%%%%%%%%%%%%%%%%%
  At large values $\xi^2\gg 1$, our analysis shows
  that the dependence on the envelope shape
  disappears because, similar to
  the Breit-Wheeler process \cite{TKTH-2013}, only
  the central part of the envelope becomes important.
  Formally, under a change of the variable
  $l\to l_{\rm eff}=l + m^2\xi^2u/2(k\cdot p)$,
  the basic functions $Y_l(z)$ with $l\gg1, \,z\gg1$
  become similar to the asymptotic
  form of the Bessel functions $J_l(z)$ and, as a consequence,
  one can get the total production probability
  (or the total cross section) in the IPA form \cite{Ritus-79}.

\subsection{Anisotropy}

\begin{figure}[h!]
 \includegraphics[width=0.85\columnwidth]{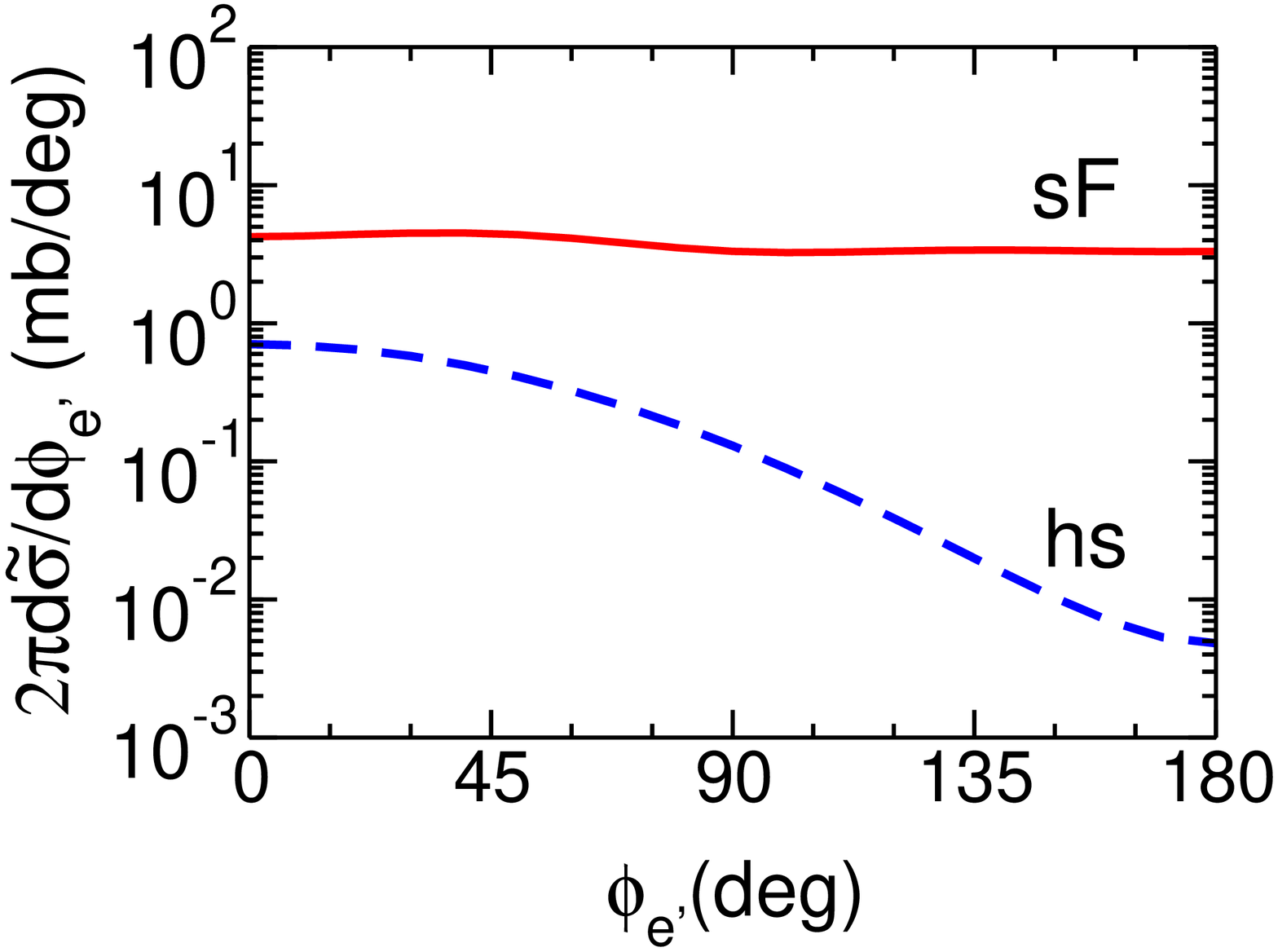}\\
 \includegraphics[width=0.85\columnwidth]{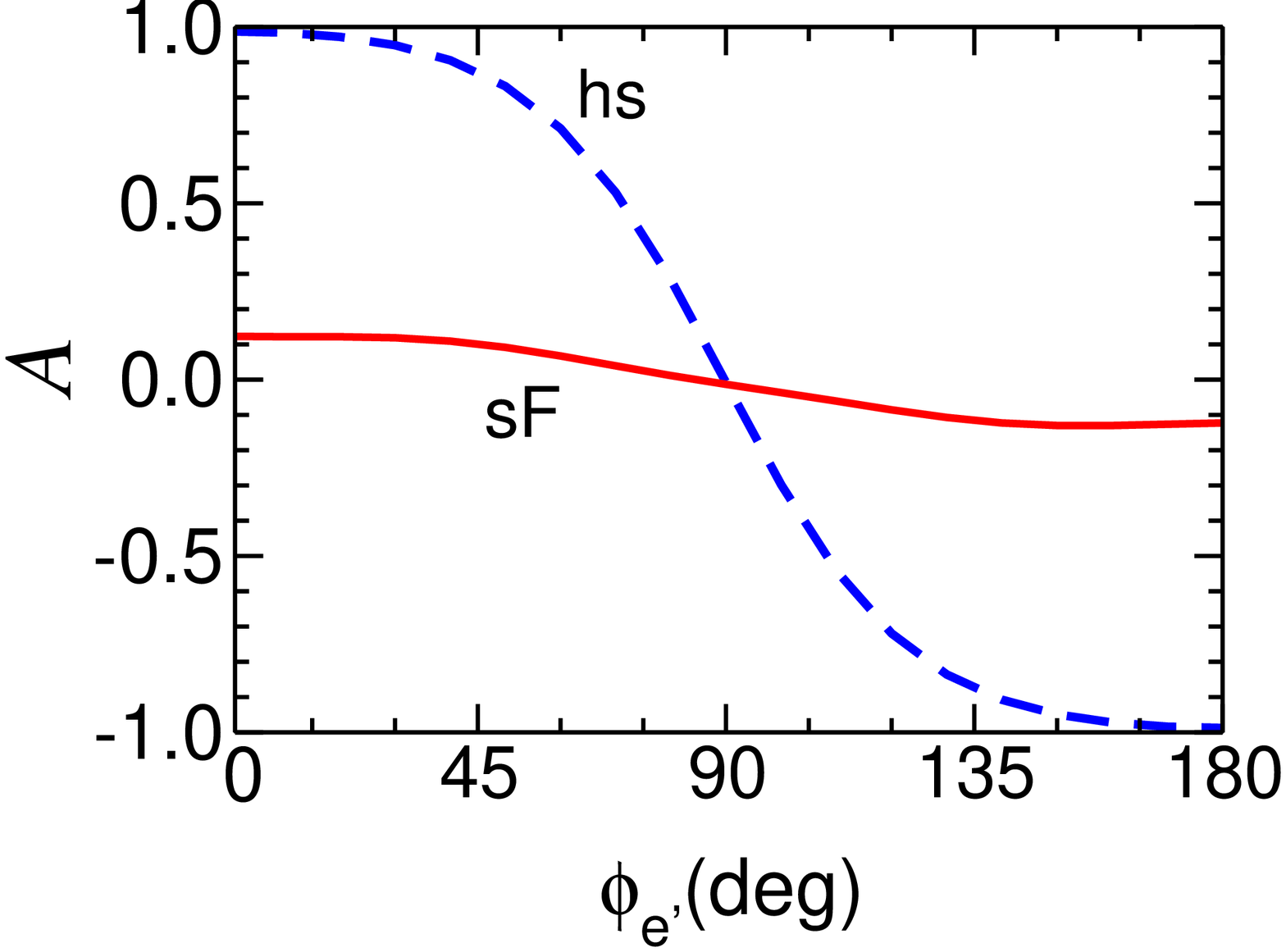}
 \caption{\small{(Color online)
 Top panel: Integrated cross sections
 as a function of azimuthal angle of the outgoing electron
 for $N=0.5$, $\xi^2=0.1$ and $\omega'/\omega'_1=3$.
 The dashed blue and solid red curves correspond to the hyperbolic secant (hs)
 symmetrized Fermi (sF), respectively.
 Bottom panel: The same as in the top panel, but for the electron anisotropy
 defined in Eq.~(\ref{S8}).
 \label{Fig:12}}}
 \end{figure}

 Let us discuss now the $\phi_{e'}$ (azimuthal angle) dependence of
 the outgoing electron for sub-cycle pulse.
 The corresponding cross sections are defined by Eqs.~(\ref{S1}) and
 (\ref{S33}) with fixed azimuthal angle $\phi_{e'}$.
% The azimuthal angle dependence is determined mostly by the
% oscillating factor $\exp[il\phi -iz {\cal P}(\phi,\phi_{e'})]$
% in the basic
% functions in~Eq.~(\ref{III24}).
 The anisotropy is manifest most clearly
 in case of the sub-cycle pulse with finite field intensity,
 and it is very sensitive to the pulse shape.
 Thus, for  sub-cycle and a smooth
 one-parameter envelope shape
 (e.g., for the hs pulse shape) and
 finite field intensity,
 the direction of flight of the outgoing
 electron (photon) is correlated with the coordinate frame of the
 e.m.\ field. The production probability would have a maximum
 if the azimuthal angle of the outgoing electron $\phi_{e'}=\phi_0$
 is equal to zero, or the azimuthal direction of the electron momentum
 coincides with $\vec a_x$.
The  explanation of this effect is as follows.
 The most important dependence of the basic functions
 $Y_l(z)$  in~(\ref{III24}),
 which determine the production probability $w(l)$ in (\ref{S2}),
 on azimuthal angle $\phi_{e'}$ is
 \begin{eqnarray}
 Y_l(z)\propto \int\limits_{-\infty}^{\infty}\,
 d\phi\,{f}(\phi)\,{\rm e}^{il\phi-iz
 \int\limits_{-\infty}^\phi d\phi'f(\phi')\cos(\phi'-\phi_{e'})}~,
 \label{ANI1}
\end{eqnarray}
 and similarly for $X_l(z)$ with the substitution $f\to f^2$ in the integrand.
 The argument in the highly oscillating exponent is
 \begin{eqnarray}
 R(\phi)&=&i\left[l\phi -z\cos\phi_{e'}
 \int\limits_{-\infty}^\phi d\phi'f(\phi')\cos\phi'\right.\nonumber\\
 &-&\left. z\sin\phi_{e'}\int\limits_{-\infty}^\phi d\phi'f(\phi')\sin\phi'
 \right].
 \label{ANI2}
 \end{eqnarray}
 The contributions of  $Y_l^2$ and $X_l^2$ to the probability would
 take a maximum when $R$ is minimal. At finite values of $l$ and $z$,
 the dominant contribution comes from small $\phi'\sim 1/l$.
 Moreover, utilizing the one-parameter sub-cycle
 fast-decreasing envelope shape leads to the inequality
\begin{eqnarray}
 \int\limits_{-\infty}^\phi d\phi'f(\phi')\cos\phi'\gg
 \int\limits_{-\infty}^\phi d\phi'f(\phi')\sin\phi'~,
 \label{ANI3}
\end{eqnarray}
 and therefore the second integral in (\ref{ANI2}) can be
 neglected. Then obviously, $R$ would have a minimum at maximum
 value of $\cos\phi_{e'}$, i.e.\ at $\phi_{e'}=0$.
 This argument does not apply to the sF envelope, where the first and
 the second integrals in (\ref{ANI2}) are of the same order
 of magnitude, and the dependence $R$ on $\phi_{e'}$ becomes
 very weak.
 These properties of the partially integrated cross sections
 are illustrated in Fig.~\ref{Fig:12}, top panel,
 where results for the hs and sF envelope shapes are shown by
 the dashed blue and solid red curves, respectively.

To quantify the anisotropy we define
\begin{eqnarray}
 {\cal A}(\phi_{e'})=
 \frac
 {\tilde\sigma(\omega',\phi_{e'})- \tilde\sigma(\omega',\pi+\phi_{e'})}
 {\tilde\sigma(\omega',\phi_{e'})+ \tilde\sigma(\omega',\pi+\phi_{e'})}~.
 \label{S8}
 \end{eqnarray}
 In Fig.~\ref{Fig:12} (bottom panel),
 the anisotropy is shown as a function
 of the azimuthal angle $\phi_{e'}$ in the multi-photon region
 for $\omega'/\omega_1'=3$ and for $\xi^2=0.1$.
  One can see a distinct
  anisotropy for the hs shape and a negligible one
  for the sF flat-top shape. Our result for the hs shape
  qualitatively coincides with the result of a recent
  paper~\cite{Seipt-2013},
  where an analog analysis is done for the one-parameter envelope shape
  $f(\phi)=\cos^2(\phi/2N)\,\theta(2N-|\phi|)$
  which is qualitatively similar to the hyperbolic secant in Eq.~(\ref{E1})
  for  $N\geq 1$. The situation changes for the
  sF envelope, where the anisotropy is very small
  for the sub-cycle pulse with $N \simeq 0.5$, as shown in
  Fig.~\ref{Fig:12}, bottom panel. Therefore, one
  can conclude that the anisotropy
  strongly depends on the envelope shape.\\

\section{Summary}

  In summary we have considered the generalized
  nonlinear (multi-photon) effects in Compton scattering of
  short and ultra-short (sub-cycle)
  laser pulses for different pulse shapes.
  Such pulses might be produced in future
  facilities. %\cite{Piazza}.
  In particular, we have shown that the fully differential
  cross sections as a function of the frequency of the
  outgoing photon at fixed production angle are rapidly
  oscillating functions for short pulses with the duration
  determined by the number of oscillations $N=2\cdots 10$,
  especially for the flat-top envelope shapes. An
  experimental study of multi-photon effects in case of rapidly
  oscillating cross sections seems to be rather challenging.
 %%%%%%%%%%%%%%%%%
  To overcome the problem of
  such a staggering we suggest to utilize the
  partly integrated  cross section
  which seems to be a powerful tool for studying
  the non-linear (multi-photon) dynamics in the sub-threshold region.
  We find that these cross sections at selected
  pulse properties (field intensity, pulse duration)
  are very sensitive to the pulse shape.
  In the case of small e.m.\ field intensities, the cross sections may be
  enhanced by several orders of magnitude as compared to
  an infinitely long pulse. Such an enhancement is more important
  for flat-top envelope shapes which generate intensive
  high-frequency harmonics and play a role of a power
  amplifier.
  In the above-threshold region, the partly integrated cross sections
  manifest some "universality", i.e. the independence of the pulse
  shape structure, where results for FPA and IPA are close to each other.
  Note that such a "universality" does not appear in fully
  differential cross sections, where one can find
  rapidly oscillating cross sections as a function of $\omega'$,
  especially for the flat-top envelope shape  (cf. Figs.~2 and 3).
  A smooth one-parameter envelope shape leads
  to a non-trivial anisotropy of the outgoing electrons
  (photons) for very short pulses.
  At high field intensity, the central part of envelopes
  becomes dominant and the integrated cross sections coincide with
  the results for infinitely long pulses.
  It provides a rationale for the use of simple analytical
  expressions of IPA for Monte Carlo transport approaches.

  In this work,  we consider processes with  circularly polarized
  photon beams. We expect that qualitatively,
  in the case of a linearly polarized  pulse, our main result,
  i.e.\ the sensitivity of partially
  integrated cross sections to the sub-threshold multi-photon
  interactions and to the pulse structure, would be similar.
  The main difference is expected for the anisotropy since
the  momentum of the outgoing electron will be correlated with the direction
of  pulse polarization. Technically, the case of linear polarization
  is more complicated because it needs an additional basic function
  $Y_{2l}$, which is an analog of the $A_2$ function in IPA
  (cf. [17]). Therefore we restrict our consideration
  to the simple and clear example of circular polarization
  and will analyze the  case of linear polarization
  in  follow-up work.

\acknowledgments
 The authors acknowledge fruitful discussions with
 R.~Sauerbrey, T.~E.~Cowan and D.~Seipt.

%%%%%%%%%%%%%%%%%%%%%%%%%%%%%%%%%%%%%

\end{document}